\documentclass[usenatbib]{mnras}
\usepackage{amsmath}
\usepackage[T1]{fontenc}

\DeclareRobustCommand{\VAN}[3]{#2}
\let\VANthebibliography\thebibliography
\def\thebibliography{\DeclareRobustCommand{\VAN}[3]{##3}\VANthebibliography}

\usepackage{graphicx}	% Including figure files
\usepackage{amssymb}	% Extra maths symbols
\usepackage{booktabs}
\usepackage[utf8]{inputenc}
\usepackage{xcolor}
\usepackage{tcolorbox}
\usepackage{hyperref}
\usepackage{cleveref}
\usepackage{appendix}
\definecolor{blue(ryb)}{rgb}{0.01, 0.28, 1.0}
\hypersetup{colorlinks,citecolor={blue(ryb)}%    citebordercolor=blue
}  

\usepackage{dsfont}

\AtBeginDocument{\hypersetup{pdfborder={0 0 0}}}% <-----------

\usepackage{kantlipsum}
\usepackage{graphicx}
\usepackage{txfonts}
\usepackage{dblfloatfix}    % To enable figures at the bottom of page
  \usepackage{subfig}

\def\equationautorefname~#1\null{eq.~(#1)\null}

\newcommand{\aref}[1]{\hyperref[#1]{Appendix~\ref*{#1}}}
\title[Halo mass function from NRE]{Towards reconstructing the halo clustering and halo mass function of N-body simulations using neural ratio estimation}

\author[A. Dimitriou et al.]{
Androniki Dimitriou,$^{1,~2,~}$\thanks{E-mail: androniki.dimitriou@ific.uv.es}
Christoph Weniger,$^{2,~}$\thanks{E-mail: c.weniger@uva.nl}
Camila A. Correa,$^{2,~}$\thanks{E-mail: c.a.correa@uva.nl}
\\
$^{1}$Instituto de F\'{i}sica Corpuscular (IFIC), Universitat de Val\'{e}ncia-CSIC, E-46980, Valencia, Spain\\
$^{2}$GRAPPA (Gravitation Astroparticle Physics Amsterdam), University of Amsterdam, Science Park 904, 1098 XH Amsterdam, The Netherlands
}

\date{Accepted XXX. Received YYY; in original form ZZZ}

\pubyear{2022}

\begin{document}
\label{firstpage}
\pagerange{\pageref{firstpage}--\pageref{lastpage}}
\maketitle

\begin{abstract}
 High-resolution cosmological N-body simulations are excellent tools for modelling the formation and clustering of dark matter haloes. These simulations suggest complex physical theories of halo formation governed by a set of effective physical parameters.  
 Our goal is to extract these parameters and their uncertainties in a Bayesian context. We make a step towards automatising this process by directly comparing dark matter density projection maps extracted from cosmological simulations, with density projections generated from an analytical halo model. The model is based on a toy implementation of two body correlation functions.
  To accomplish this we use marginal neural ratio estimation, an algorithm for simulation-based inference that allows marginal posteriors to be estimated by approximating marginal likelihood-to-evidence ratios with a neural network. In this case, we train a neural network with mock images to identify the correct values of the physical parameters that produced a given image.
 Using the trained neural network on cosmological N-body simulation images we are able to reconstruct the halo mass function, to generate mock images similar to the N-body simulation images and to identify the lowest mass of the haloes of those images, provided that they have the same clustering with our training data.
 Our results indicate that this is a promising approach in the path towards developing cosmological simulations assisted by neural networks.
  
\end{abstract}


%%%%%%%%%%%%%%%%%%%%%%%%%%%%%%%%%%%%%%%%%%%%%%%%%%

%%%%%%%%%%%%%%%%% BODY OF PAPER %%%%%%%%%%%%%%%%%%

\section{Introduction}
\label{introduction}
High-precision galaxy surveys, such as Sloan Digital Sky Survey (SDSS; \citealt{2000AJ....120.1579Y}; \citealt{2009ApJS..182..543A}), the Dark Energy Survey (DES; \citealt{2016MNRAS.460.1270D}), the Dark Energy Spectroscopic Instrument (DESI; \citealt{2016arXiv161100036D}), EUCLID (\citealt{2021arXiv210801201S}), and the upcoming Legacy Survey of Space and Time (LSST; \citealt{2009arXiv0912.0201L}; \citealt{2019ApJ...873..111I}), allow the detailed study of large scale structures and galaxy clustering statistics.

Interpreting the galaxy distribution in these surveys demands accurate theoretical predictions. A historical approach in the analysis of galaxy clustering lies in the analytic modelling of the formation and evolution of the dark matter haloes (\citealt{1974ApJ...187..425P}; \citealt{1991ApJ...379..440B}; \citealt{1996MNRAS.282..347M}; \citealt{Sheth_1999}).
Even though this is one of the most active fields of astrophysics, it remains very challenging because of its multi-scale and multi-physics character. Cosmological numerical simulations (see e.g., \citealt{2005Natur.435..629S}; \citealt{2020NatRP...2...42V}) are the method of choice for tackling the growing complexities at small scales. In these simulations dark matter builds the backbone for structure formation. Dark matter particles track the influence of gravity and are able to accurately reproduce clustering on small scales. Once we have access to these simulations, we identify real objects in them: dark matter haloes and subhaloes (\citealt{Springel_2001}; \citealt{2013ApJ...770...57B}) with which galaxies can be linked.

In the past few years, projects such as Eagle (\citealt{2015MNRAS.446..521S}), Horizon-AGN (\citealt{10.1093/mnras/stw2265}), MassiveBlack-II (\citealt{10.1093/mnras/stv627}), the Illustris simulations (\citealt{Vogelsberger_2014}; \citealt{10.1093/mnras/stu1713}; \citealt{10.1093/mnras/stx3112}), among others, have demonstrated that simulations of structure formation can reproduce many small-scale structural properties such as galaxy morphology and luminosity, galaxy stellar mass function and star formation rates of galaxies. These simulations are computationally expensive to produce, especially when baryonic physics and increasing resolution are included.  To circumvent this problem, and exploit the wealth of data produced by cosmological simulations, machine learning has become a powerful tool (see e.g.,  \citealt{Lucie_Smith_2018}; \citealt{2019MNRAS.484.5771A}; \citealt{2019MNRAS.482.2861B};  \citealt{2020arXiv201200240A};  \citealt{2020PhRvD.101l3525A}; \citealt{Ntampaka_2020}).

With the improvement in image recognition, classification and segmentation (e.g. \citealt{2015arXiv150504597R}), machine learning techniques have been applied in a large number of astronomical studies related to galaxy properties (e.g. \citealt{Dieleman_2015}; \citealt{article1}; \citealt{Hocking_2017}; \citealt{2020MNRAS.491.2506D}; \citealt{2021arXiv211101154B}). In the context of halo modelling, machine learning algorithms have been implemented to predict galaxy properties based on halo formation and evolution (\citealt{Xu_2013}; \citealt{2018MNRAS.478.3410A}; \citealt{https://doi.org/10.48550/arxiv.2012.00111}; \citealt{Lovell_2021}; \citealt{2021MNRAS.504.4024M}; \citealt{Xu_2021}).

\medskip

In this work, our goal is to measure the halo population parameters from a set of dark-matter-only cosmological simulations. Those simulations can be also thought of as parametric stochastic simulators, that take as input a set of physical parameters, sample many latent nuisance variables, and finally produce simulation data.
%what is known as the inverse problem. 
%We are
%interested in calculating  cosmological simulations the casual factors that produced them in the bayesian setting, i.e, we want to marginalize over the physical  parameters of our theory. Despite
%of this being a universal problem in data analysis, it is very difficult to solve. The reason is that the multidimensional nature of the data suggests complicated models, which can be also thought as parametric stochastic simulators, that have as input a large number of physical parameters as well as nuisance latent variables.

The applicability of likelihood based inference techniques, which are the golden standard to solve parameter inference problems in many areas, % such as like Markov chain Monte Carlo (MCMC) (\cite{1953JChPh..21.1087M},  \cite{10.1093/biomet/57.1.97}) or nested sampling \citep{10.1214/06-BA127}, 
is severely limited in a high-dimensional setting,
since they can be very time consuming, or inference is even impossible as the large number of parameters often leads to intractable likelihood functions.
During the past years, likelihood-free inference techniques, also known as simulation-based inference,  (\citealt{cranmer2020frontier}) have become state-of-the-art in solving various inverse problems by training neural networks (NN) to perform the mapping from observations to physical parameters.
The main idea behind these techniques is to try to approximate posteriors instead of sampling from them. Likelihood-free inference is applied in several astrophysical contexts (see e.g., \citealt{2017JCAP...05..037B}; \citealt{2018MNRAS.477.2874A}; \citealt{2019MNRAS.488.4440A}; \citealt{2019ApJ...886...49B}; \citealt{2019ApJ...886...49B}; \citealt{2020arXiv201012931D}; \citealt{2020arXiv201007032C}; \citealt{cole2021fast}; \citealt{2021PhRvL.127x1103D}; \citealt{2021MNRAS.507.1999H}; \citealt{2021JCAP...11..049M}; \citealt{2021arXiv210909747V}; \citealt{2021arXiv210910360V}; \citealt{2022ApJ...926..151Z}; \citealt{2022MNRAS.512..661K}; \citealt{2022MNRAS.511.3046H} and \citealt{2022arXiv220509126A}).

 Here, we construct an analytical forward toy model/simulator that generates images depicting the surface density of clustered dark matter haloes. We directly calibrate the model on dark-matter-only cosmological simulations using  Marginal Neural Ratio Estimation (MNRE; \citealt{miller2020simulationefficient}; \citealt{Miller:2021hys}). This model contains four physical parameters: the number of haloes, $N$, the slope of the halo mass function, $a$, and two parameters of an effective clustering model, $\epsilon$ and  $n$, as well as  many latent nuisance variables which also control the spatial distribution of the haloes and are related to the halo mass function. We look for the best fit parameters on dark-matter-only simulations in order to reconstruct the halo mass function, to identify the lowest mass of the haloes in those simulations and to generate images which look similar to them.
 
\medskip

 The paper is structured as follows.
 In \autoref{method}, we explain the structure of our forward model and we define its parameters. We also describe the process of its calibration on dark-matter-only simulation data generated by the EAGLE project (\citealt{2015MNRAS.446..521S}), using MNRE. 
 In \autoref{results}, we analyze the results that we obtain, focusing on reconstructing the halo mass function, as well as an artificial lower cutoff that we add at it.
 Finally, in \autoref{conclusion}, we summarize our results, we spot the limitations of our model and we discuss directions of future improvement.

\begin{figure*}
\hspace{0.2cm}
\begin{minipage}{0.33\textwidth}
\subfloat[\label {0-0}]{\includegraphics[scale=.7]{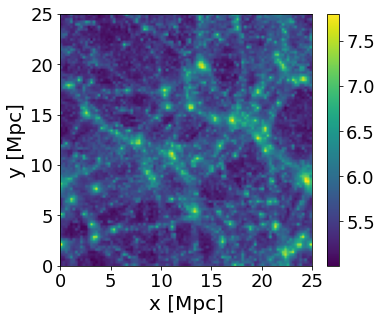}}
\end{minipage}%
\hspace{-0.2cm}
\begin{minipage}{0.33\textwidth}
\subfloat[\label {0-1}]{\includegraphics[scale=.7]{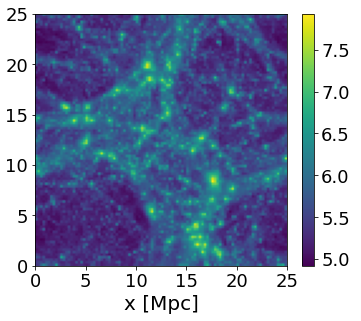}}
\end{minipage}
\hspace{-0.4cm}
\begin{minipage}{0.33\textwidth}
\subfloat[\label {0-2}]{\includegraphics[scale=.7]{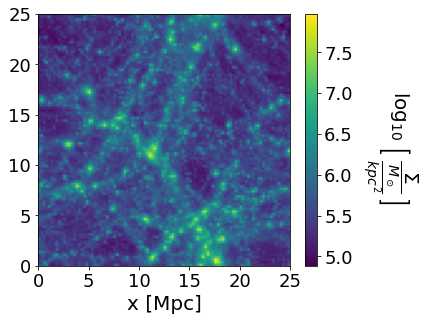}}
\end{minipage}
\caption{Logarithmic surface density of dark matter in units of $\frac{\text{M}_\odot}{\text{kpc}^{2}}$, calculated from two-dimensional histograms of the redshift-zero snapshot of a dark matter-only (25 Mpc$)^{3}$ simulation. In \autoref{0-0}, coordinate z $\in$ (0,25) Mpc is projected, while in \autoref{0-1} and \autoref{0-2} coordinates y $\in$ (0,25) Mpc and x $\in$ (0,25) Mpc are projected.}
\label{fig0}
\end{figure*}

\section{Method}
\label{method}

Our goal is to reconstruct the halo clustering and the halo mass function of dark-matter-only cosmological simulations, by directly calibrating an analytical toy halo model on them. The simulations that we use are described in \autoref{simulations}, while a full description of our model and of the methodology that we follow for the calibration is presented in \autoref{toy halo model} and \autoref{simulation based inference} respectively.

\subsection{N-body simulation data}
\label{simulations}
In this work, we use the dark matter-only cosmological simulations generated by the EAGLE project \citep{2015MNRAS.446..521S}. More specifically, we focus on the L025N0376 simulation, which corresponds to a comoving volume of 25 $\text{Mpc}$ on a side with $376^{3}$ dark matter particles, each one having a mass of  $1.15\cdot 10^{7}  $ $\text{M}_{\odot}$.
Our toy halo model generates images that depict the surface density of clustered dark matter haloes. Since we want to compare these images with dark matter-only cosmological simulations images, we generate projected density maps of the 25 $\text{Mpc}$ cosmological box. We do that by using the coordinates of the dark matter particles provided by the redshift zero simulation snapshot. Examples of these maps in units of $\frac{\text{M}_{\odot}}{\text{kpc}^{2}}$ and can be seen in \autoref{fig0}.

%Although the filaments observed in these images play a crucial role in galaxy evolution, very little is known about their number density and distribution and therefore it is very difficult to reproduce them through an analytical model. 

Although the filaments observed in these images play a crucial role in galaxy evolution, very little is known about their number density and distribution and therefore it is very difficult to reproduce them through an analytical model. Thus, we suppress them by focusing on dark matter haloes with  masses inside a specific range. For the masses we use the definition of virial mass, ${M}_{200\rho_{crit}}$, which is the total mass  contained  within the virial radius of the halo, where the mean density is equal to 200 times the critical density of the universe at that time. In addition, we select only the dark matter particles that are gravitationally bound to them.  We also further suppress filamentary structures by looking at slices of the simulation box with reduced thickness, $\Delta z$.

\begin{figure*}
\begin{minipage}{0.33\textwidth}
\subfloat[\label {1-0}]{\includegraphics[scale=.43]{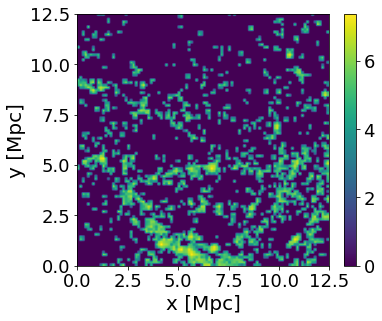}}
\end{minipage}%
\begin{minipage}{0.33\textwidth}
\subfloat[\label {1-1}]{\includegraphics[scale=.7]{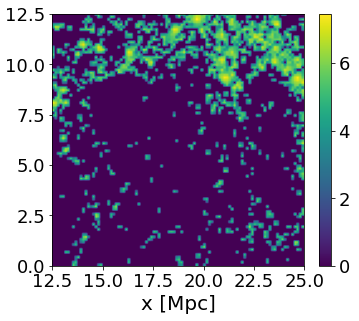}}
\end{minipage}
\hspace{-0.5cm}
\begin{minipage}{0.33\textwidth}
\subfloat[\label {1-2}]{\includegraphics[scale=.7]{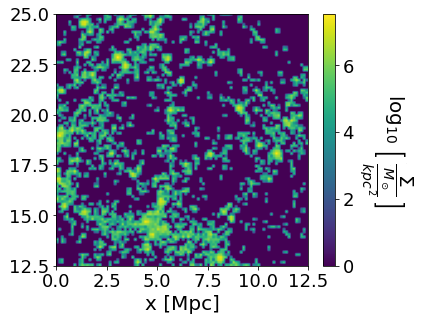}}
\end{minipage}
\caption{Logarithmic surface density of dark matter haloes in units of $\frac{\text{M}_\odot}{\text{kpc}^{2}}$, calculated from two-dimensional histograms of the redshift zero snapshot of the (25 Mpc$)^{3}$ simulation. These densities correspond to different 12.5x12.5x12.5 $\text{Mpc}^{3}$ sub-boxes of the simulation and to haloes with mass $M_{200\rho_{crit}} \in (10^{9},10^{12})$ $\text{M}_{\odot}$.
In all images coordinate $z \in (0,12.5)$ $\text{Mpc}$ is projected.}
\label{fig1}
\end{figure*}

Finally, in order to generate more images from the L025N0376 simulation, we look at projected density maps of different sub-volumes of the 25 $\text{Mpc}$ cosmological box. Using the coordinates of the centers of potential (CoP) of the haloes, we select only those which belong to $12.5\times12.5\times \Delta z$ $\text{Mpc}^{3}$ subvolumes, where \mbox{$\Delta z \in$ (8, 25) Mpc}. For example, by choosing haloes with masses ${M}_{200\rho_{crit}} \in(10^{9},10^{12})$ ${\text{M}}_{\odot}$ and CoP's inside three different \mbox{$12.5\times12.5\times12.5$ $ \text{Mpc}^{3}$} boxes of the L025N0376 simulation, we get different projections, in which the filamentary structure is suppressed (see \autoref{fig1}). 

%The way to do that is by utilizing both the EAGLE project's snapshot and its FoF halo finder. Using the mass definitions that the halo finder provides, we  first choose haloes with the desired masses and then we read the group numbers of these haloes. 

\subsection{Toy halo model}
\label{toy halo model}
Next, we build a simple forward model consisting of four physical parameters, that generates images similar to the ones shown in \autoref{fig1}. 
As mentioned in \autoref{simulations}, those images are projected density maps of different sub-volumes of the 25Mpc cosmological box. 

Therefore, the first parameter of our model is the number of haloes, $N$, that we will place on a projected plane. 
We sample halo masses from a halo mass function of the form
\begin{align}
    \frac{dn}{dM}=b\left(\frac{M}{{\text{M}_\odot}}\right)^{-a},
    \label{halomassfunction}
\end{align}
(see e.g., \citealt{2001MNRAS.321..372J}; \citealt{2006}; \citealt{2007}; \citealt{2008}; \citealt{2009}) where $a$, the slope of the halo mass function, is the second parameter of our model. As further described in \aref{sampling}, the sampling procedure does not depend on the normalization, $b$. Thus, $b$ is not a parameter of our model.

Once we obtain the masses of the $N$ haloes, we calculate their concentrations using the concentration-mass relation derived by \cite{Correa_2015b}.

\subsubsection{Spatial distribution}
\label{spatialdistribution}
Next, we place the haloes on a projected plane. Since haloes in actual N-body simulation images are clustered, we mimic this behaviour by introducing correlations amongst the positions of different halos. The most simple way to do it, in terms of computational speed and efficiency, is 
by using two-point correlation functions (or their Fourier transforms, power spectrums) and Gaussian fields, which are completely determined, in a statistical sense, by them.
Therefore, we  sample the positions of the haloes according to distributions generated from 2D realizations of  Gaussian random fields on an $M\times M$ grid.
In our case, the Gaussian field, $\delta^{\textbf{x}}_{ab}$, is specified by a power spectrum given by a power-law
\begin{align}
    P(k)\sim\frac{1}{k^{n}},
    \label{powersp}
\end{align}
where $n$ is the slope of the power spectrum, which is the third parameter of our model (for more details see \aref{method_clustering} and \aref{mock data application}).

\begin{figure*}
\begin{minipage}{0.32\textwidth}
\subfloat[$N$=1626, $a$=1.62, $\epsilon$=1,  $n$=0.9\label {4-0}]{\includegraphics[scale=.43]{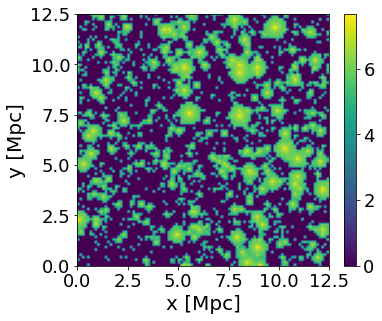}}
\end{minipage}%
\hspace{0.2cm}
\begin{minipage}{0.32\textwidth}
\subfloat[$N$=826, $a$=2.4, $\epsilon$=1.2,  $n$=5\label {4-1}]{\includegraphics[scale=.7]{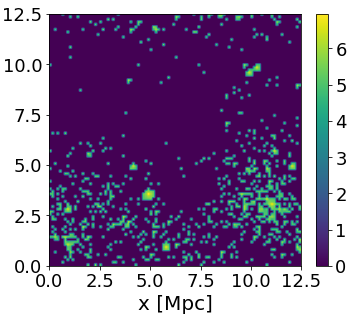}}
\end{minipage}
\hspace{-0.3cm}
\begin{minipage}{0.32\textwidth}
\subfloat[$N$=2026, $a$=1.78, $\epsilon$=1,  $n$=7 \label {4-2}]{\includegraphics[scale=.7]{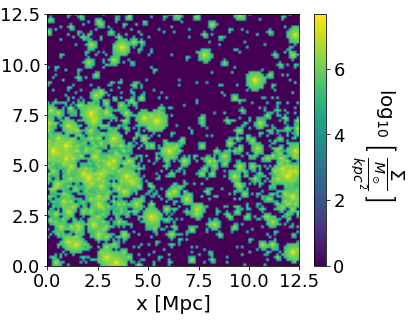}}
\end{minipage}
\caption{Logarithmic surface density of dark matter haloes with masses $\text{M}_{h}\in (10^{9},10^{12})$ ${\text{M}}_{\odot}$, in units of $\frac{\text{M}_\odot}{\text{kpc}^{2}}$. These images are mock data generated by the simple toy halo model described in \autoref{toy halo model}, which contains four physical parameters and includes clustering. In each panel the parameters, i.e, the number of dark matter haloes ($N$), the slope of the halo mass function ($a$) and the parameters that control the clustering, namely the exponent multiplied with density field ($\epsilon$) and the slope of the power spectrum ($n$), are randomly generated.}
\label{fig4}
\end{figure*}

Once we have the Gaussian field, $\delta^{\textbf{x}}_{ab}$, we convert it into a probability distribution function, in order to be able to sample the positions of the haloes from it. 
To do that, we first multiply $\delta^{\textbf{x}}_{ab}$ with a parameter $\epsilon$, which is the fourth parameter of our model. We calculate the field $f=e^{\epsilon\cdot \delta^{\textbf{x}}_{ab}}$ to get strictly positive values, and we normalize it so that its values sum to one. The role of the parameter $\epsilon$ is to further control the over and under densities in the Gaussian field produced by a given power spectrum. For example, $\epsilon$=0 will result in a uniform distribution, while a large $\epsilon$ will generate pronounced  peaks. However, the parameter $\epsilon$ and the field $\delta^{\textbf{x}}_{ab}$ can be very correlated or degenerate. In order to reduce these effects, before exponentiating, we normalize the field  $\delta^{\textbf{x}}_{ab}$ such as to have a constant variance, which is set to one.

Then, we use the normalized field $f$ to sample the positions of the haloes, and finally, we place each halo in the 2D plane. What we obtain are  images where the values of their pixels correspond to the distance between the center of the pixel and a fixed reference which is the position of the halo, i.e., we obtain the projected radius, $r'$, of each halo. More details about the sampling of the positions of the haloes and the calculation of their projected radius, $r'$, can be found in \aref{details position}.

Since we know the projected radius, $r'$, the mass, $M_{h}$, and the concentration, $c$, of each halo, we can use them to calculate the halo's surface density. We assume that haloes follow an NFW density profile \citep{Navarro_1997}.
The surface density of these haloes, in units of $\frac{\text{M}_{\odot}}{\text{kpc}^{2}}$ and in terms of the projected radius, $r'$, the scale radius, $r_{s}$ and the concentration, $c$, is equal to
    
\begin{align}
    \rho_{2D}(r')&= \rho^{0}_{c}\cdot \delta_{char} \cdot f_{2D}(r') \nonumber \\
    &=277.5 h^{2}\cdot \frac{200 c^{3}g(c)}{3} \cdot f_{2D}(r'),
    \end{align}
    where
    \begin{align}
g(c)=\frac{1}{\ln(1+c)-c/(1+c)},
\end{align}
%\text{ \citep{1996}}
   and $f_{2D}(r')$ is the 2D projection of the NFW profile
   \begin{align}
   \rho(r)=\frac{\rho_{c}^{0}\delta_{char}}{\left(\frac{r}{r_{s}}\right)\left(1+\frac{r}{r_{s}}\right)^{2}},
   \label{eq_NFW}
   \end{align}
which is given by
\begin{align}
f_{2D}(r')=-\frac{2r_{s}^{3}\cdot\left[r'\cdot(r_{s}^{2} - r'^{2}) + 2r_{s}\cdot r'\cdot \arctan(\frac{r'\cdot \sqrt{r' - r_{s}}}{r'\cdot \sqrt{r_{s} + r'}})\cdot  \sqrt{r'^{2} - r_{s}^{2}}\right]}{r'\cdot (r_{s}^{4} - 2r_{s}^{2}\cdot r'^{2} + r'^{4})}.
\label{density}
\end{align}
This formula is derived analytically in \cite{androniki} and is equivalent to the one presented in \cite{2001MNRAS.321..155L}.
Using the definition of the halo concentration

\begin{align}
    c\equiv \frac{r_{h}}{r_{s}},
\end{align}
and of the virial mass of a halo which follows an NFW profile,
\begin{align}
    M_{h}=\frac{4\pi}{3}200\rho^{0}_{c}r^{3}_{h},
    \label{mh}
\end{align}
we can express \autoref{density} in terms of the mass, $M_{h}$, of the halo and its concentration, $c$, since
\begin{align}
    r_{s}=\frac{r_{h}}{c}=\frac{(\frac{3M_{h}}{4\pi\rho^{0}_{c}200})^{1/3}}{c}=\frac{(3M_{h})^{1/3}}{(800\pi\rho^{0}_{c})^{1/3}c}.
    \label{scale}
\end{align}

After calculating the surface density of a halo at all pixels, we set the values of the pixels that correspond to projected radii larger than $2.7\cdot r_{h}$ equal to 1, where $r_{h}$ is the halo's virial radius. %, defined as the radius within which the mean density is equal to 200 times the critical density of the Universe at that time. 
In this way, we define an artificial boundary for each halo which depends on its mass, $M_{h}$. Once we calculate the $\log_{10}$ of the pixels of these images, the regions after that boundary will be empty. We do that in order to be able to construct images similar to those of \autoref{fig1}.

By adding images of multiple haloes together, we obtain the total surface density field, generated by the $N$ haloes. Constructing the grids in the way that we described above, results in an image which has the same dimensions and pixel size as the ones of \autoref{fig1}. 
    
Finally, we add poisson noise to the image, since this is also the noise that actual N-body simulations have, due to their  finite number of particles. The noisy image is obtained by sampling each of its pixels from a poisson distribution with $\lambda$ equal to the value of the corresponding pixel of the noiseless image.
Examples of images (mock data) generated from different combinations of the parameters of our model can be seen in \autoref{fig4}.

\subsection{Simulation-based inference}
\label{simulation based inference}
As described above, our forward model consists of a set of parameters, $\boldsymbol{z}$. The parameters $\boldsymbol{z}$ are either physical parameters, denoted by $\boldsymbol{\theta}$, or latent nuisance variables denoted by $\boldsymbol{\omega}$, i.e., we have $\boldsymbol{z}=\boldsymbol{\theta}\cup\boldsymbol{\omega}$.  %takes as input a set of parameters $\boldsymbol{z}$ and produces data $\boldsymbol{x}\sim p(\boldsymbol{x}|\boldsymbol{z})$. These parameters are both physical as well as latent nuisance variables.  
The physical parameters are the number of haloes, $N$, the slope of the halo mass function, $a$, the exponent, $\epsilon$, and the slope of the power spectrum, $n$. The latent nuisance variables are all the variables that are generated randomly in different parts of the model.
Thus, more specifically, our model  takes as input a vector of parameters $\boldsymbol{\theta}$, samples many
internal states  $\boldsymbol{\omega}$, and finally
produces a data vector $\boldsymbol{x}\sim p(\boldsymbol{x}|\boldsymbol{z})=p(\boldsymbol{x}|\boldsymbol{\theta},\boldsymbol{\omega})$. 

Given the observed data $\boldsymbol{x}$, we want to infer the input parameters $\boldsymbol{\theta}$.
Obtaining marginal posteriors of the parameters of interest using likelihood-based inference methods, involves evaluating the likelihood function $p(\boldsymbol{x}|\boldsymbol{\theta}$). However, this likelihood is defined implicitly by the forward model and often is not tractable, since it should be calculated by the high-dimensional integral  

\begin{align}
    p(\boldsymbol{x}|\boldsymbol{\theta}) =\int d\boldsymbol{\omega}  p(\boldsymbol{x,\omega}|\boldsymbol{\theta}),
\end{align}
where $p(\boldsymbol{x,\omega}|\boldsymbol{\theta})$ is the joint probability density of data $\boldsymbol{x}$ and  of the very large latent space $\boldsymbol{\omega}$ \citep{cranmer2020frontier}. 
%Thus, Bayesian inference as well as marginalisation over the physical parameters of interest, cannot be performed using likelihood-based inference methods. 

\subsubsection{Neural Ratio Estimation (NRE)}
\label{ratioestimation}
One recently proposed likelihood-free inference technique is Neural Ratio Estimation (NRE) (\citealt{hermans2020likelihoodfree}). The goal of Ratio Estimation is to approximate the ratio
\begin{align}
    r(\boldsymbol{x},\boldsymbol{\theta})\equiv \frac{p(\boldsymbol{x},\boldsymbol{\theta})}{p(\boldsymbol{x})p(\boldsymbol{\theta})}=\frac{p(\boldsymbol{x}|\boldsymbol{\theta})}{p(\boldsymbol{x})}= \frac{p(\boldsymbol{\theta}|\boldsymbol{x})}{p(\boldsymbol{\theta})}.
    \label{ratio}
\end{align}
Given the prior $p(\boldsymbol{\theta})$ of the parameters, learning the ratio is enough in order to estimate the posterior $p(\boldsymbol{\theta}|\boldsymbol{x})$.
According to \cite{hermans2020likelihoodfree} it is possible to estimate this ratio by training a neural network (NN) as a parametrized binary classifier $d(\boldsymbol{x},\boldsymbol{\theta})$ to distinguish dependent sample-parameter pairs $(\boldsymbol{x},\boldsymbol{\theta})\sim p(\boldsymbol{x},\boldsymbol{\theta})$ with class label $y=1$ from independent sample-parameter pairs $(\boldsymbol{x},\boldsymbol{\theta})\sim p(\boldsymbol{x})p(\boldsymbol{\theta})$ with class label $y=0$. 

We would like the output of the NN $d_{\phi}(\boldsymbol{x},\boldsymbol{\theta})$, where $\phi$ are the network weights, to be interpreted as the probability of class with \mbox{label 1},
\begin{align}
   d_{\phi}(\boldsymbol{x},\boldsymbol{\theta})&=p(y=1|\boldsymbol{x},\boldsymbol{\theta})= \\
   &=\frac{p(\boldsymbol{x},\boldsymbol{\theta}|y=1)p(y=1)}{p(\boldsymbol{x},\boldsymbol{\theta}|y=1)p(y=1)+p(\boldsymbol{x},\boldsymbol{\theta}|y=0)p(y=0)}= \nonumber \\
   &=\frac{p(\boldsymbol{x},\boldsymbol{\theta})}{p(\boldsymbol{x},\boldsymbol{\theta})+p(\boldsymbol{x})p(\boldsymbol{y})},
\label{function}
\end{align}
assuming equal class population.
Since we deal with a binary classifier, the binary-cross entropy loss function can be used in order to train the network via stochastic gradient descent. In the limit of infinite training data the loss functional is
\vspace{0.2cm}

\noindent $ L[d_{\phi}(\boldsymbol{x},\boldsymbol{\theta})]$
\begin{equation}
   \resizebox{\columnwidth}{!}{$=\int d\boldsymbol{\theta} \int d\boldsymbol{x} p(\boldsymbol{x},\boldsymbol{\theta})[-\log d_{\phi}(\boldsymbol{x},\boldsymbol{\theta})]+p(\boldsymbol{\theta})p(\boldsymbol{x})[-\log(1-d_{\phi}(\boldsymbol{x},\boldsymbol{\theta}))]$}.
\end{equation}
This loss functional is minimized exactly for the function $d_{\phi}(x,\theta)$ defined at \autoref{function}, as described in \aref{proof}.

After training the NN with a given number of training data, its output will be  
\begin{align}
    d(\boldsymbol{x},\boldsymbol{\theta})\approx\frac{p(\boldsymbol{x},\boldsymbol{\theta})}{p(\boldsymbol{x},\boldsymbol{\theta})+p(\boldsymbol{x})p(\boldsymbol{\theta})}.
      \label{d}
\end{align}
By using \autoref{d}, we can estimate directly the posterior
$p(\boldsymbol{\theta}|\boldsymbol{x})$ as
\begin{align}
    p(\boldsymbol{\theta}|\boldsymbol{x})\approx \frac{d(\boldsymbol{x},\boldsymbol{\theta})}{d(\boldsymbol{x},\boldsymbol{\theta})-1}p(\boldsymbol{\theta})=\hat{p}(\boldsymbol{\theta}|\boldsymbol{x}),
\end{align}
(for more details see \aref{posterior estimation}).

\subsubsection{Marginal Neural Ratio Estimation (MNRE)}
\label{sec:swyft}

In order to obtain the marginal posteriors of the physical parameters, we will apply Marginal Neural Ratio Estimation (MNRE; \citealt{Miller:2021hys}). MNRE is implemented via \textit{swyft}\footnote{https://github.com/undark-lab/swyft} (\citealt{miller2020simulationefficient}), an open-source software package, which estimates marginal likelihood-to-evidence ratios, that are amortized over a range of observational data $\boldsymbol{x}$ and physical parameters $\boldsymbol\theta$.

According to this method, first we define the parameters of interest  $\boldsymbol{\vartheta_{k}}$. The parameters $\boldsymbol{\vartheta}_{k}$ include physical parameters of the model associated with all the one dimensional marginal posteriors such as \{$\theta_{1}$, $\theta_{2}$, $\theta_{3}$, $\theta_{4}$\}=\{$N$, $a$, $\epsilon$, $n$\}, parameters associated with all two dimensional marginal posteriors such as $\{(\theta_{i},\theta_{j})\in \mathbb{R}^{2}|i=1,...4,j=i+1,...4\}$ and unions of these sets. 

\textit{Swyft} then uses NRE to estimate the corresponding marginal likelihood-to-evidence ratios
\begin{align}
    r_{k}(\boldsymbol{x},\boldsymbol{\vartheta}_{k}):=\frac{p(\boldsymbol{x}|\boldsymbol{\vartheta_{k}})}{p(\boldsymbol{x})}=\frac{p(\boldsymbol{\vartheta}_{k}|\boldsymbol{x})}{p(\boldsymbol{\vartheta}_{k})}.
    \label{ratiomargswyft}
\end{align}
More details can be found in \aref{MNRE_appendix}.
\begin{figure*}
\centering
\includegraphics[width=18cm,height=10cm,keepaspectratio]{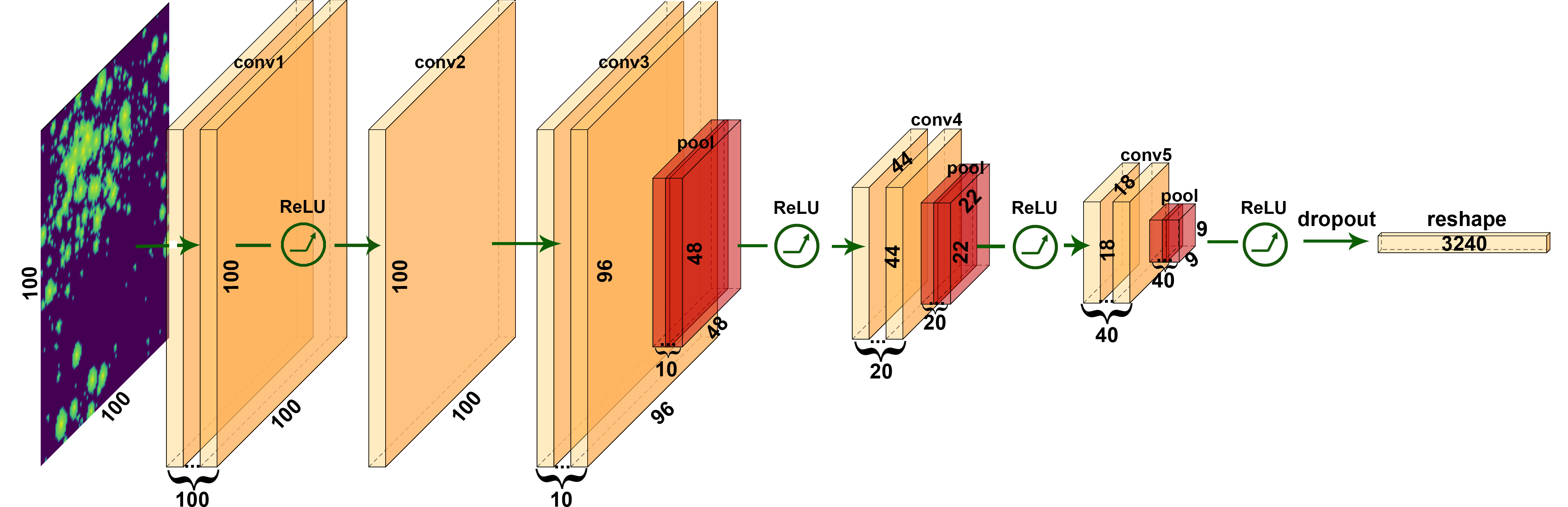}
\caption{The CNN used to preprocess the mock data produced by the toy halo model described in \autoref{toy halo model}. The CNN consists of 13 layers: conv1 $\rightarrow$ ReLU $\rightarrow$ conv2 $\rightarrow$ conv3 $\rightarrow$ max-pool  $\rightarrow$ ReLU $\rightarrow$ conv4 $\rightarrow$ max-pool  $\rightarrow$ ReLU $\rightarrow$ conv5 $\rightarrow$ max-pool  $\rightarrow$ ReLU  $\rightarrow$ dropout. After the final dropout layer, the data are reshaped into a 1D array which is the input for an MLP, as described in \autoref{sec:swyft}.}
\label{fig5}
\end{figure*}

\begin{figure*}
%\hspace{-0.3cm}
\begin{minipage}{0.32\textwidth}
\vspace{0.2cm}
\subfloat[\label {888}]{\includegraphics[scale=.6]{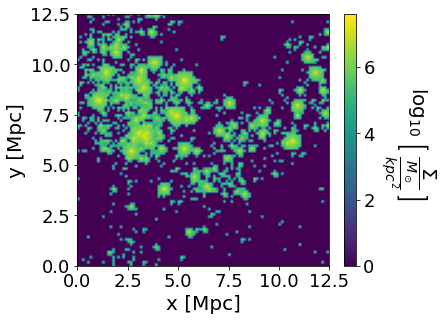}}
\end{minipage}%
\begin{minipage}{0.68\textwidth}
\subfloat[ \label {999}]{\includegraphics[scale=.52]{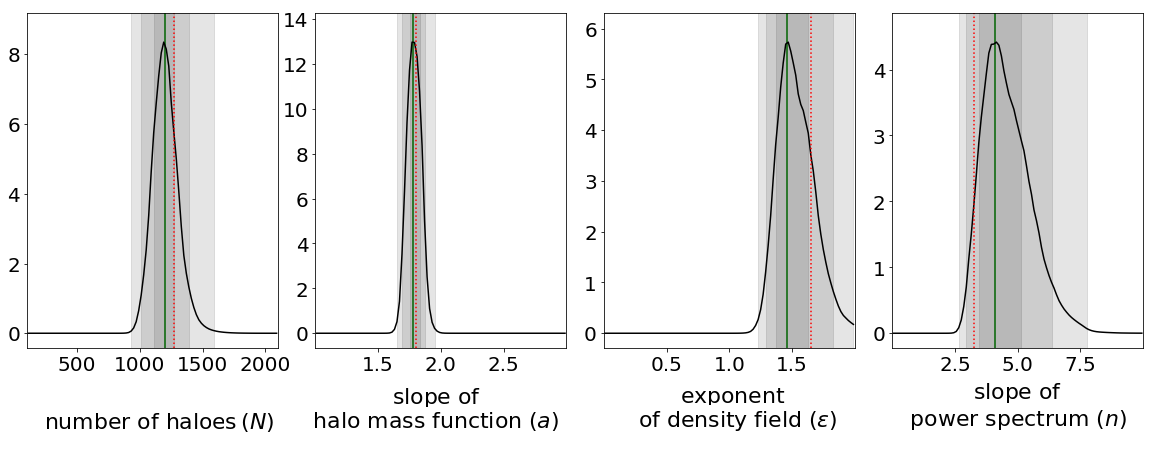}}
\end{minipage}
\caption{\autoref{888} is generated from the toy halo model described in \autoref{toy halo model}, using  $N$=1253, $a$=1.78, $\epsilon$=1.64 and  $n$=3.21 and depicts the logarithmic surface density of dark matter haloes with masses $M_{h}\in (10^{9},10^{12})$ $\text{M}_{\odot}$, in units of $\frac{\text{M}_\odot}{\text{kpc}^{2}}$. \autoref{999} illustrates the one dimensional marginal posteriors for the four physical parameters of the toy halo model. The correct values of the parameters (vertical red dashed lines), the modes of the posteriors (vertical green lines) and their 68.72\%, 95.45\% and 99.73\% highest posterior density regions (grey regions) can be seen.}
\label{figmock_new}
\end{figure*}

%\begin{figure*}
%\centering
%\begin{minipage}{1\linewidth}
%\centering
%\subfloat[]{\includegraphics[scale=.45]{images/mock1.png}}
%\end{minipage}\par\medskip
%\caption{One dimensional marginal posteriors of the parameters $N$, $a$, $\epsilon$ and $n$, corresponding to the image (\ref{1-0}) and a rotation of it. The modes (green dots), the 68.72, 95.45 and 99.73\% HDIs of the posteriors and the correct values of parameters $N$ and $a$ (red dashed lines) are also plotted.}
%\label{fig12}
%\end{figure*}

\begin{figure*}
\hspace{-0.5cm}
\begin{minipage}{0.24\textwidth}
\subfloat[\label{666}]{\includegraphics[scale=.52]{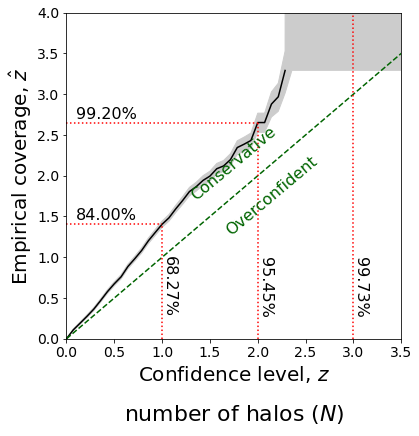}}
\end{minipage}
\hspace{0.1cm}
\begin{minipage}{0.24\textwidth}
\subfloat[]{\includegraphics[scale=.52]{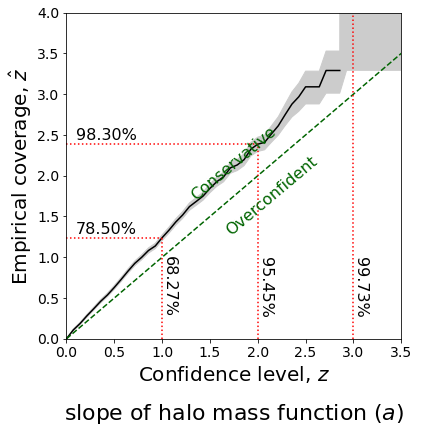}}
\end{minipage} 
\hspace{0.1cm}
\begin{minipage}{0.24\textwidth}
\subfloat[]{\includegraphics[scale=.52]{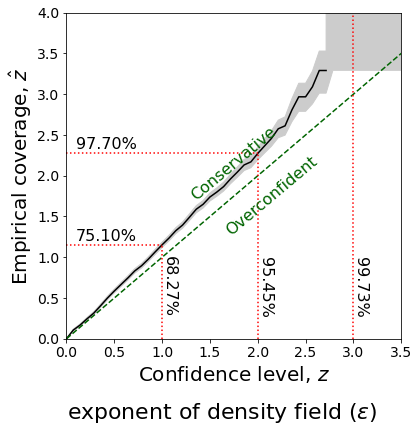}}
\end{minipage}
\hspace{0.1cm}
\begin{minipage}{0.24\textwidth}
\subfloat[]{\includegraphics[scale=.52]{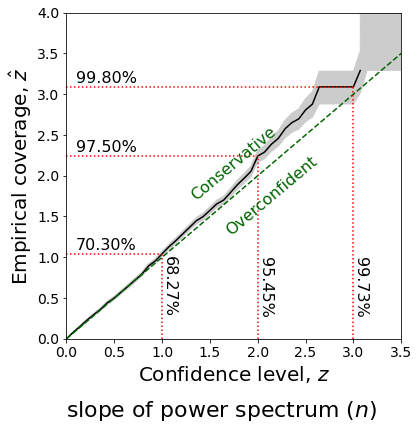}}
\end{minipage}

\caption{Empirical expected coverage probability  as function of the confidence level, 1 - $\alpha$, or alternatively of $z$, for the four physical parameters of the toy halo model described in \autoref{toy halo model}. The probabilities are calculated using simulation data. In cases where $\hat{z}$ > z, the estimated credible intervals are conservative and contain the true
values with a frequency higher than nominally expected. The nominal probabilities 1 - $\alpha$ are shown
in vertical dashed red lines, while the empirical estimates 1 - $\hat{\alpha}$ are shown in the horizontal ones.}
\label{fig66}
\end{figure*}

\subsection{Training}
\label{training}
In order to start training binary classifiers to get marginal posteriors of the physical parameters of our model, we first define the priors of the parameters and the number of training data.
These parameters, together with their priors, can be seen in \autoref{table1}.

%\begin{table}[!b]
%\centering 
%\resizebox{\columnwidth}{!}{%
%\begin{tabular}{||c c c||}
 %\hline
%  Parameter & Prior   & Range\\ 
% \hline\hline
%$N$: Number of haloes & Uniform & 100-2000  \\ 
% \hline
%$a$: Inner slope of halo mass function & Uniform  &1-3 \\
% \hline
%$\epsilon$: Exponent of the gaussian random field & Uniform  %&0-2\\
% \hline
%$n$: Slope of the power spectrum  & Uniform  &0-10 \\
% \hline
%\end{tabular}}
% \caption{The parameters of the toy halo model together with the priors used during training.} \label{table1}
%\end{table}

Moreover, we choose to train with 200.000 mock data samples and we define a preprocessor network that acts as feature extractor: a Convolutional Neural Network (CNN), whose architecture can be seen in \autoref{fig5}.
We train using a batch size of 64, a learning rate of $10^{-4}$ and for a maximum of 100 epochs.

\begin{table}
\captionsetup{size=footnotesize}
\caption{The physical parameters of the toy halo model described in \autoref{toy halo model} together with their uniform priors used during training.} \label{tab:freq}
\setlength\tabcolsep{0pt} % let LaTeX compute intercolumn whitespace
\centering 

\smallskip 
\begin{tabular*}{\columnwidth}{@{\extracolsep{\fill}}lcc}
\toprule
Parameter   & Range\\ 
\midrule
 $N$: Number of haloes & 100-2000  \\ 
$a$: Inner slope of halo mass function   &1-3 \\
$\epsilon$: Exponent of the Gaussian random field  &0-2\\
$n$: Slope of the power spectrum  & 0-10 \\
\bottomrule
\end{tabular*}
\label{table1}
\end{table}

\section{Results}
\label{results}

This Section is structured as follows.
In \autoref{mockresults} we evaluate the performance of the trained NN on mock data. In \autoref{evalnbody}, we test its performance on N-body simulation data. More specifically, in \autoref{reconhmf} we focus on reconstructing the halo mass function, while in \autoref{lowercutoff} on estimating the lowest mass of the haloes, or alternatively, an artificial lowest cutoff of the halo mass function of the N-body simulation images.

\subsection{Inference of simulation model}

\label{mockresults}

Examples of the resulting one dimensional posteriors of the parameters of two mock images, together with their correct values and their highest posterior density regions, can be seen in \autoref{figmock_new}. The figure shows that the NN is able to reconstruct the four parameters since their correct values (vertical dashed lines) lie inside the $1\sigma$ and $2\sigma$ intervals of their corresponding posteriors.

To test the performance of the trained NN, and therefore the quality of the posteriors and their credible regions,  we use mock data and the notions of the nominal and empirical expected coverage probabilities.
  Following \cite{hermans2021averting}, the empirical expected coverage probability of the 1 - $\alpha$ highest posterior density regions, derived from the posterior estimator $\hat{p}(\boldsymbol{\vartheta_{k} | x})$, given
a set of $n$ i.i.d. mock data ($\boldsymbol{\vartheta^{*}}_{k i}$ , $\boldsymbol{x}_{i}) \sim p(\boldsymbol{\vartheta}_{k}, \boldsymbol{x})$, can be calculated as

\begin{align}
   1-\hat{\alpha}= \frac{1}{n}\sum_{i=1}^{n}\mathds{1}[\boldsymbol{\vartheta^{*}_{k i}} \in \Theta_{\hat{p}(\boldsymbol{\vartheta_{k}|x_{i}})}(1-\alpha)].
\end{align}

In this expression $\Theta_{\hat{p}(\boldsymbol{\vartheta_{k}}|\boldsymbol{x})}$(1 - $\alpha)$ is the 1 - $\alpha$ highest posterior density region of $\hat{p}(\boldsymbol{\vartheta_{k}}|\boldsymbol{x})$ and $\boldsymbol{\vartheta}^{*}_{k}$  is the ground truth  specified to the simulation model, whose forward evaluation produced the observable $\boldsymbol{x}$.
On the other hand, the nominal expected coverage probability is the expected coverage probability of the true posterior and is equal to the confidence level $1-\alpha$. Therefore, this expression allows us to estimate the actual error rate $\hat{\alpha}$ of a $1 - \alpha$ highest posterior density region of $\hat{p}(\boldsymbol{\vartheta_{k}}|\boldsymbol{x})$.
In the case of a perfectly calibrated posterior the empirical expected coverage probability is equal to the nominal expected coverage probability. This means that when we randomly generate observations $(\boldsymbol{x}$, $\boldsymbol{\vartheta_{k}}) \sim p(\boldsymbol{x | \vartheta_{k}})p(\boldsymbol{\vartheta_{k}})$, we expect the true parameters $\boldsymbol{\vartheta^{*}_{k}}$ to fall outside of the 1-$\alpha$ region in $\alpha$ of the cases (\citealt{cole2021fast}).
According to \cite{hermans2021averting}, a posterior estimator is considered to be acceptable whenever the empirical expected coverage probability is larger or equal to the nominal expected coverage probability.

\autoref{fig66} highlights our results. In these panels we reparametrize $\alpha$ ($\hat{\alpha}$) with a new variable $z$ ($\hat{z}$), defined as the 1-$\frac{ 1}{2}\alpha$ (1-$\frac{ 1}{2}\hat{\alpha}$) quantile of the
standard normal distribution, following \cite{cole2021fast}. Therefore, the most commonly used “1$\sigma$”, “2$\sigma$” and “3$\sigma$” regions that correspond to 1- $\alpha$ = 0.6827, 0.9545, 0.9997 are quoted as  z = 1, 2, 3. Furthermore, we
denote the empirical coverage  in terms of 1 - $\hat{\alpha}$ (horizontal numbers) for the above mentioned confidence levels (vertical numbers).
Finally, we estimate uncertainties of the expected coverage probability due to the finite total number of mock data, $n$, using Jeffreys interval (\citealt{cole2021fast}).

By looking at \autoref{fig66}, we see that the 68.27\% (95.45\%) highest
posterior density region corresponding to parameter $N$, contains the ground truth value in 84.00\% (99.20\%) of the cases, and is hence conservative, or in other words the confidence intervals over-cover. In the case of perfect coverage, one would expect the black and the green dashed
line to perfectly overlap.
 Moreover, we observe that the highest posterior density regions corresponding to parameters $a$, $\epsilon$ and $n$, are slightly conservative too. Therefore, following \cite{hermans2021averting}, our posterior estimators are conservative.

%while the empirical expected coverage probability of the 1 - $\alpha$ highest posterior density region (HPDR) of some estimated posterior $\hat{p}(\boldsymbol{\vartheta_{k}}|\boldsymbol{x})$ is then given by:
%\begin{align}
%    \displaystyle \mathop{\mathbb{E}_{p(\boldsymbol{\vartheta,x)}}}[(\mathds{1}[\boldsymbol{\vartheta_{k}} \in \Theta_{\hat{p}(\boldsymbol{\vartheta_{k}|x})}(1-\alpha)]]
%\end{align}

%For example, a
%95\% highest posterior density region would have an expected error rate of α = 0.05. In other words,
%when randomly drawing observations x, θ ∼ p(x | θ)p(θ), we expect the true parameters θ to fall
%outside of the 95% region in 5% of the cases.

%we start by feeding it with mock data, i.e., with images that we generate with our model by specifying the values of its four parameters. Examples of the resulting one dimensional marginal posteriors of the parameters, together with their correct values, can be seen in Figures (\ref{mock:kako}), (\ref{mock:kako1}) and  (\ref{mock:kako2}). 

%\clearpage

\subsection{N-body simulation images}
\label{evalnbody}
Now, we can test the NN's performance on actual N-body simulation images, like those of \autoref{fig1}. At first, we will test if the NN is able to reconstruct the halo mass function.

\subsubsection{Reconstructing the halo mass function}
\label{reconhmf}
As opposed to mock images, in the case of N-body simulations, only two of the correct values of the parameters are known: the number, $N$, of the haloes in each image, which can be calculated by using the data provided by the FOF halo finder and the slope, $a$, of the  halo mass function, which is found to be $\sim$1.9 (\citealt{zavala2019dark}). Thus, we would like to see whether the NN is able to reconstruct the halo mass function, i.e., if it is able to identify the correct values of $N$ and $a$ and also to predict the values of the clustering parameters $\epsilon$ and $n$  for sub-boxes of the simulation L025N0376, like those of \autoref{fig1}. To increase the size of our test data set we also rotate each sub-box by 90, 180 and 270 degrees. The highest posterior density regions that correspond to 16 sub-boxes, together with the correct values of the parameters $N$ and $a$, can be seen in \autoref{fig12}.

\begin{figure}
\centering
\begin{minipage}{0.32\textwidth}
\subfloat[\label{661}]{\includegraphics[scale=0.54]{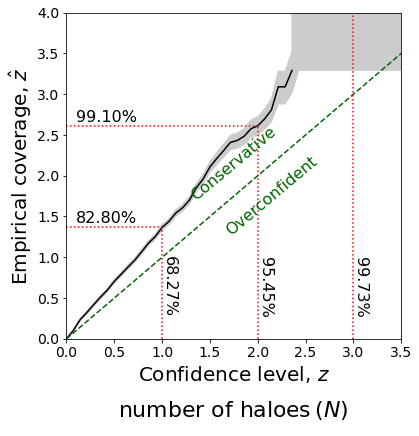}}
\end{minipage} \\
\begin{minipage}{0.32\textwidth}
\subfloat[]{\includegraphics[scale=0.55]{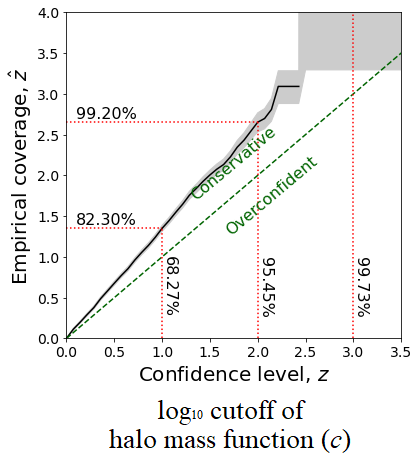}}
\end{minipage} 
\caption{Like \autoref{fig66}, but for the parameters $N$ and $c$ of the toy halo model described in \autoref{lowercutoff}.}
\label{figcovtes2}
\end{figure}

\begin{figure*}
\centering
\includegraphics[width=15cm,height=15cm,keepaspectratio]{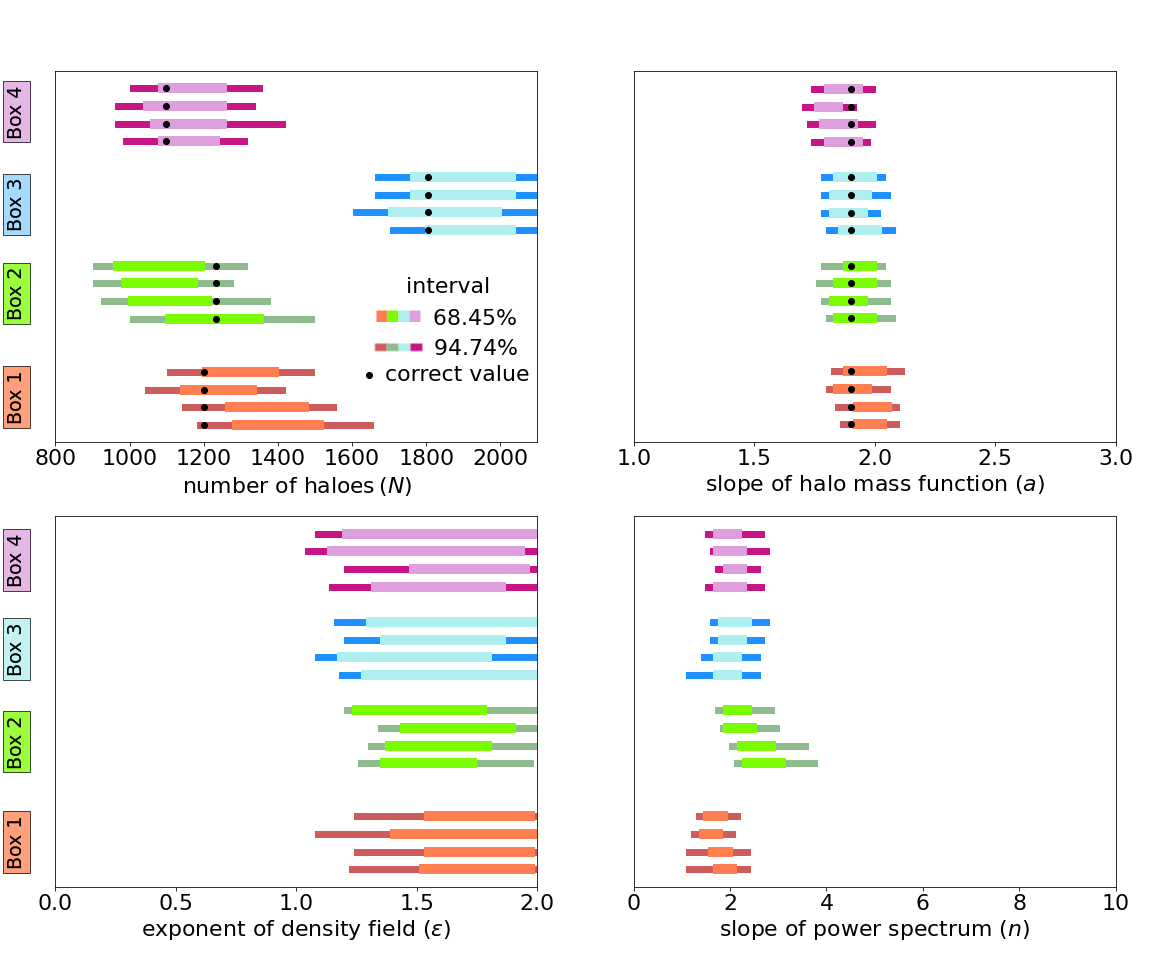}
\caption{(Main Result 1) 68.45\% and 94.74 \% highest posterior density regions of the four physical parameters of a simple toy halo model that includes clustering, which is directly derived from dark-matter-only simulations using Marginal Neural Ratio Estimation (MNRE). Each color corresponds to results obtained by a sub-box of a (25 Mpc$)^{3}$ simulation and three rotations of it, for which we know the correct values for the number of haloes, $N$, and the slope of halo mass function, $a$.
We reproduce the number of haloes and the slope of halo mass function reasonably well (the correct values lie inside the $1\sigma$ and $2\sigma$ intervals), while we predict the values of $\epsilon$ and $n$ which are effective parameters of a clustering model that is based on a toy implementation of two-body correlation functions. We also observe that the results obtained by each sub-box and the three rotations are very stable, despite the fact that rotations change the images completely. }
\label{fig12}
\end{figure*}

%An inference network trained on a simple toy model applied to real data

%\begin{figure*}
%\hspace{-1cm}
%\begin{minipage}{.5\linewidth}
%\subfloat[\label{cont1}]{\includegraphics[%scale=.48]{images/contours.png}}
%\end{minipage}
%\hspace{1cm}
%\begin{minipage}{.5\linewidth}
%\vspace{0.1cm}
%\subfloat[\label{cont2}]{\includegraphics[%scale=.48]{images/corner_plot_2params.png}}
%\end{minipage}
%\caption{One and two-dimensional posteriors of the physical parameters of the two toy halo models described in \ref{toy halo model} and \ref{lowercutoff}. Fig. (\ref{cont1}) is obtained by image (\ref{1-1}), while Fig. (\ref{cont2}) by image (\ref{1-2}). The contours for the 68.45\%, 94.74\% and 99.73\% highest posterior density regions, along with the correct values of the parameters $N$, $a$ and $c$ (red lines), are shown. We observe that the parameters $\epsilon$ and $n$ are degenerate. On the contrary, Fig. (\ref{cont2}) illustrates a negative correlation between the parameters $N$ and $c$. }
%\label{figcontours}
%\end{figure*}

\begin{figure*}
\centering
\begin{minipage}{.33\linewidth}
\subfloat[]{\includegraphics[scale=.58]{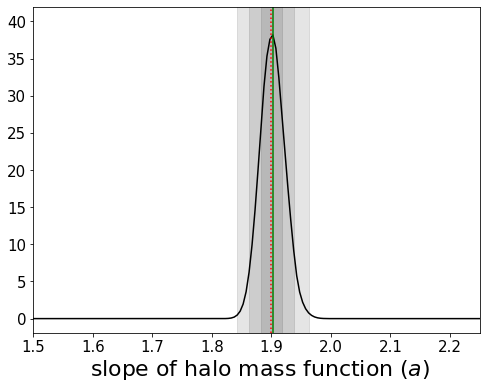}}
\end{minipage}%
\begin{minipage}{.33\linewidth}
\subfloat[]{\includegraphics[scale=.58]{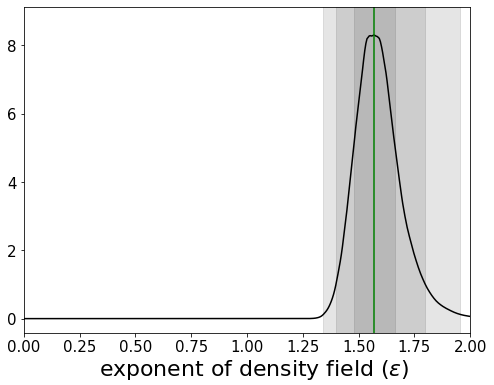}}
\end{minipage}
\begin{minipage}{.33\linewidth}
\subfloat[]{\includegraphics[scale=.58]{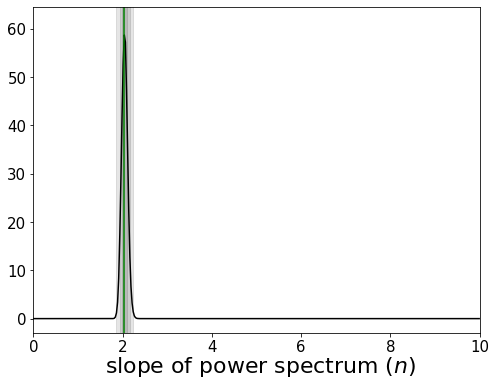}}
\end{minipage}

\caption{Combined posteriors of the parameters $a$, $\epsilon$ and $n$ of the toy halo model described in \autoref{toy halo model}, along with their modes (green lines), their 68.45\%, 94.74\% and 99.73\% highest posterior density regions and the correct value of the parameter $a$ (red dashed line). The combined posterior of each parameter is calculated using the results of \autoref{fig12} which are obtained by a set of independent and identically distributed N-body simulation images, according to \autoref{eq-comb}. We manage to reconstruct the slope of the halo mass function since its correct value, 1.9,  lies inside the 1$\sigma$ interval of the combined posterior.}
\label{fig16}
\end{figure*}

By looking at this figure, we see that the NN estimates correctly the number of haloes of all images, since the correct values of $N$ lie either inside the 68.45\% (1$\sigma$) or the 94.74\% (2$\sigma$)  highest posterior density regions. The same holds also for the inner slope, $a$, of the halo mass function. Moreover, we see that the posteriors of the slope, $n$, of the power spectrum are narrow, while the posteriors for the parameter $\epsilon$ are wide. 

%To observe the correlation between different parameters we can plot their two-dimensional posteriors. An example can be seen in Fig. (\ref{cont1}) where these posteriors are derived for the image input shown in Fig. (\ref{0-0}). We observe that there is a degeneracy between the parameters $\epsilon$ and $n$ which can be understood from the fact that both parameters control the peaks of the Gaussian field, as we already saw in \ref{spatialdistribution}.

Since the posteriors of $\epsilon$ are wide and they overlap, i.e., $\epsilon$ is not very constrained, we can calculate a combined posterior. We can do the same also for $a$ and $n$, since their posteriors are narrow and do not vary a lot between the different images. The combined posteriors of the above mentioned parameters are calculated using the 16 N-body simulation images, which we treat as a set of independent and identically distributed data (i.d.d) $\mathcal{X} = \{x_{1}, . . . , x_{16}\}$\footnote{In fact, not all images are exactly independent from each other. As already mentioned, some of them are obtained by rotating others by 90, 180 and 270 degrees.}. Following \cite{hermans2020likelihoodfree}, these posteriors are equal to
\begin{align}
p_{comb}({\vartheta}|\mathcal{X})&=\frac{p({\vartheta})\prod_{x \in \mathcal{X}}p(x|{{\vartheta}})}{\int p({\vartheta})\prod_{x \in \mathcal{X}}p(x|{{\vartheta}}) d{{\vartheta}}} \nonumber \\
&\sim \frac{p({\vartheta})\prod_{x \in \mathcal{X}}\hat{r}(x,{{\vartheta}})}{\int p({\vartheta})\prod_{x \in \mathcal{X}}\hat{r}(x,{{\vartheta}}) d{{\vartheta}}},
\label{eq-comb}
\end{align}
where   ${\vartheta} \in \{a,\epsilon,n\}$.
The combined posteriors of the parameters $a$, $\epsilon$ and $n$ that we obtain can be seen in \autoref{fig16}. 

Below we will generate mock images by letting the parameter $N$ vary, while we will set the values of $a$, $\epsilon$ and $n$ equal to the modes of these combined posteriors. A comparison between those mock images and images of \autoref{fig1} in pixel space, can be found in \aref{comp1}.

\subsubsection{Reconstructing the lower cutoff of the halo mass function}
\label{lowercutoff}
Now, we will test if we can identify the lowest mass of the haloes in the N-body simulation images. To this end, we first need to modify the model described in \autoref{toy halo model}.

Up to now, we were always sampling haloes with masses $M_{h}\in(10^{9},10^{12})$ $\text{M}_{\odot}$ according to the PDF of the halo mass function (more details about this can be found in \aref{sampling}). In order to be able to sample haloes with masses  $M_{h}\in(10^{c},10^{12})$ $\text{M}_{\odot}$, where $c$ is a new parameter, we will replace that PDF, with a new one defined as
\begin{align}
    \text{PDF}_{c}(M)= \frac{ (1-a)}{(10^{12})^{1-a}-(10^c)^{1-a}} \cdot \left(\frac{M}{{\text{M}_\odot}}\right)^{-a}.
    \label{PDF'}
\end{align}
In this way we add an artificial lower cutoff in the halo mass function of \autoref{halomassfunction}.
Moreover, we will utilize the results of \autoref{reconhmf} and therefore we will set the values of the parameters $a$, $\epsilon$ and $n$ equal to the modes of their combined posteriors, i.e., $a$=1.9025, $\epsilon=1.5675$ and $n$=2.0375.
Thus the new model has only two parameters: the number, $N$, of haloes and the cutoff, $c$, of the halo mass function.

\begin{figure*}
%\hspace{0.1cm}
\begin{minipage}{0.33\textwidth}
\subfloat[$c$=8.75\label {20-1}]{\includegraphics[scale=.4]{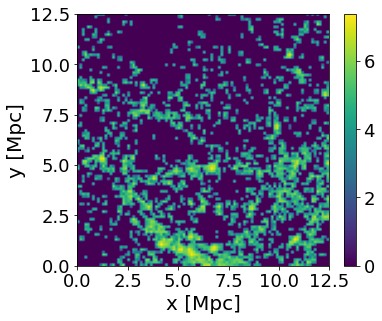}}
\end{minipage}
\begin{minipage}{0.33\textwidth}
\subfloat[$c$=9\label {20-0}]{\includegraphics[scale=.4]{images/03.png}}
\end{minipage}%
%\hspace{0.3cm}
\begin{minipage}{0.33\textwidth}
\subfloat[$c$=9.5\label {20-2}]{\includegraphics[scale=.4]{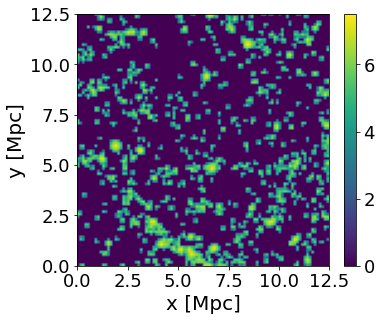}} 
\end{minipage}
\begin{center}
%    This is a mid-figure caption 
\end{center}
\caption{Logarithmic surface densities of dark matter haloes in units of $\frac{\text{M}_\odot}{\text{kpc}^{2}}$, calculated from two-dimensional histograms of the redshift zero snapshot of a (25 Mpc$)^{3}$ simulation. All images correspond to a single sub-box of the simulation, but they are generated using a different $\log_{10}$ cutoff, $c$, of the halo mass function. The left panel shows the distribution of haloes more massive than $10^{8.75}$  $\text{M}_{\odot}$, whereas the middle and right panels show the distribution of haloes more massive than $10^{9}$ and $10^{9.5}$  $\text{M}_{\odot}$ respectively.}
\label{fig20}
\end{figure*}

After defining the new model, we train a NN as before, but since we have only two parameters, we train with 50.000 mock data.
The parameters, together with their priors are shown in \autoref{paramspriors}.
\begin{table}
\captionsetup{size=footnotesize}
\caption{The physical parameters of the toy halo model described in \autoref{lowercutoff} together with the uniform priors used during training.} \label{tab:freq2}
\setlength\tabcolsep{0pt} % let LaTeX compute intercolumn whitespace
\centering 

\smallskip 
\begin{tabular*}{\columnwidth}{@{\extracolsep{\fill}}lcc}
\toprule
Parameter   & Range\\ 
\midrule
$N$: Number of haloes & 100-3000  \\ 
$c$: $\log_{10}$ lower cutoff of halo mass function &8-10.5 \\

\bottomrule
\end{tabular*}
\label{paramspriors}
\label{table2}
\end{table}
To ensure that the trained NN performs well, we perform a coverage test similar to that of \autoref{mockresults}. By looking at \autoref{figcovtes2}, we see that the 68.27\% and 95.45\% highest posterior density regions corresponding to both parameters $N$ and $c$,  are slightly conservative. Since our posterior estimators are conservative under the simulation model, we can test the NN's performance on actual N-body simulation images, like those of \autoref{fig20}. These images correspond to $\log_{10}$ surface densities produced by haloes with masses $M_{h}\in(10^{c},10^{12})$  $\text{M}_{\odot}$ where $c\in \{8.75,9,9.5\}$ $\text{M}_{\odot}$. We choose to generate images in these mass ranges, since smaller cutoffs, e.g. $c$=8 $\text{M}_{\odot}$, are inside the resolution limit of the simulations, while larger cutoffs, e.g. $c$=10 $\text{M}_{\odot}$, result in images with very few haloes.

%The results that be obtain can be seen in Figure (\ref{fig22}).  In most cases, the NN is  able to recognize the correct values of both parameters, since they always lie inside the 99.73\% highest posterior density regions. However, by looking at Box 1 with a large $\log_{10}$ cutoff, $c$, which corresponds to an image with few haloes that are not so clustered (see Figure \ref{20-2}), we see that the NN overestimates the number of haloes, $N$. On the contrary, the NN underestimates the number of haloes for Box 2 with a small cutoff, which corresponds to an image with many haloes that are very clustered (see Figure \ref{20-1}) .

\begin{figure*}
\vspace{-0.5cm}
\centering
\includegraphics[width=18cm,height=18cm,keepaspectratio]{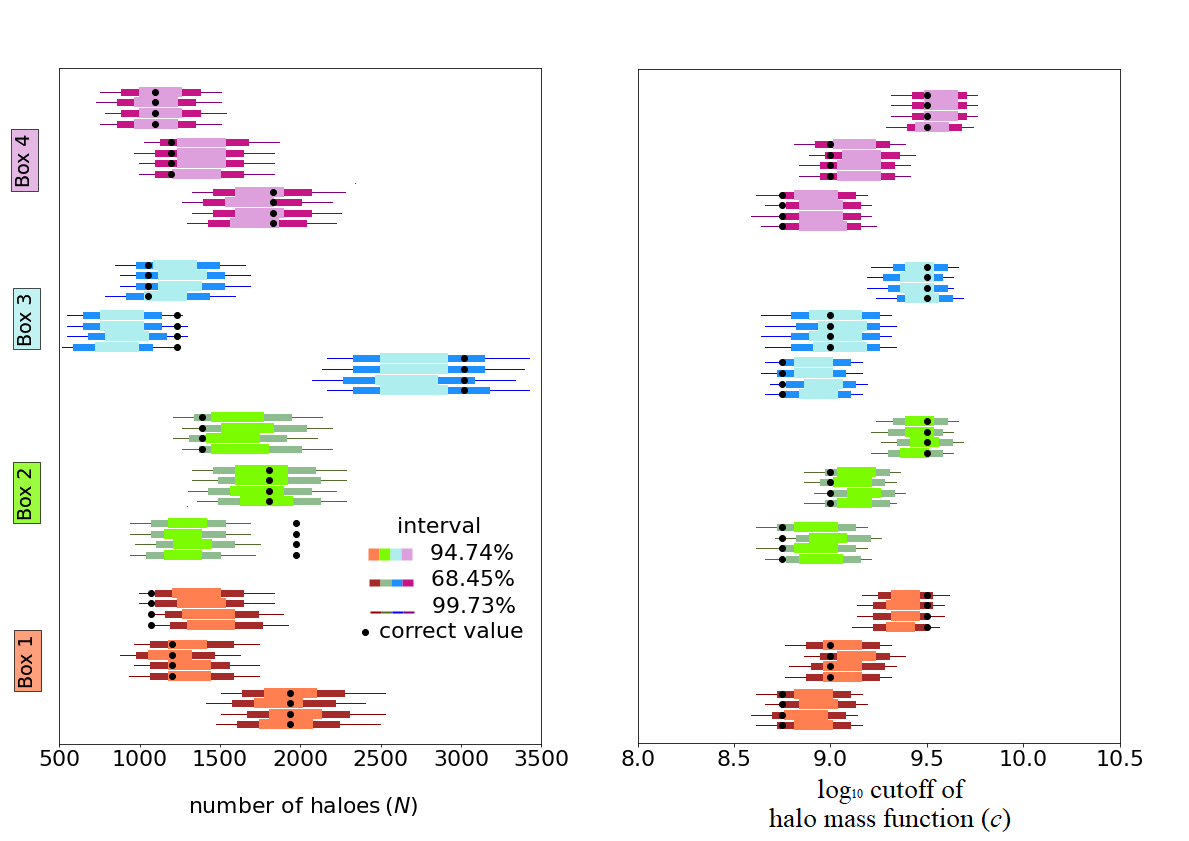}

\caption{(Main Result 2) 68.45\%, 94.74 \% and 99.73\% highest posterior density regions of the two physical parameters of the toy halo model described in \autoref{lowercutoff}, which is directly calibrated on dark-matter-only simulations using Marginal Neural Ratio Estimation (MNRE). Each color corresponds to a sub-box of a (25 Mpc$)^{3}$ simulation and 3 rotations of it. To construct this model we first use the results shown in \autoref{fig12} and we set the values of the three out of the four parameters ($a$, $\epsilon$ and $n$) of the simple toy halo model of \autoref{toy halo model}, equal to the modes of their combined posteriors shown in \autoref{fig16}. We also add a new parameter, $c$, which controls the lowest mass of the haloes in the simulations, i.e., it acts as a lower cutoff of the halo mass function.
We manage to reproduce the logarithmic cutoff reasonably well since the correct values almost always belong in the $1\sigma$ and $2\sigma$ intervals. While in most cases the same holds for the number of haloes, we observe some deficiencies, especially when looking at images with many haloes that are very clustered (see e.g., Box 2 with $c$=8.75 $\text{M}_{\odot}$). As described in \autoref{lowercutoff}, these deficiencies result from the definition of the clustering in the toy halo model. Finally, we observe again that the results obtained by each sub-box and the three rotations are very stable, despite the fact that rotations change the images completely.}
\label{fig22}
\end{figure*}

The results that be obtain can be seen in \autoref{fig22}.  We see that the NN is able to identify the correct values of the $\log_{10}$ cutoff, $c$, since they always lie inside the $1\sigma$ or $2\sigma$ intervals. In most cases, the same holds for the number of haloes, $N$, while sometimes our results are problematic, i.e., the $1\sigma$ or $2\sigma$ intervals do not contain the correct value of the parameter,  especially when looking at images with many haloes that are very clustered, e.g., at Box 2 with $c=8.75$ $\text{M}_{\odot}$. The image that corresponds to the same Box with $c=9$ $\text{M}_{\odot}$  can be seen in \autoref{1-1}. By reducing the cutoff, the number of haloes increases and all of them are accumulated in the same region. 
In this case, we see that the NN underestimates the number of haloes. This behaviour can be explained by the fact that the new model has only two parameters, while the values of the parameters $\epsilon$ and $n$, which controlled the clustering at the four-parameter model, were set equal to the modes of their combined posteriors. These modes were calculated from 16 different images that had a specific amount of clustering. %and only four of them had lots of clustering, i.e., larger values for $\epsilon$ and $n$.Thus the values of $\epsilon$ and $n$ are not so large and our training data do not contain many images with this amount of clustering. 
Therefore, our training data does not contain many images with few haloes which are almost uniformly distributed or images with many haloes that are all accumulated to a specific region.
%Moreover, we observe that the NN can more easily estimate the number of haloes from images where the cutoff is larger than 8.75. This result can be also explained in the context of clustering and of the parameters $\epsilon$ and $n$. By comparing the images of Figure (\ref{fig1}) which we used to calculate the combined posterior of $\epsilon$ and $n$, with those of Figure (\ref{fig20}), we see that even though the latter have more haloes, most of them are accumulated in certain regions. Thus, the images look similar to each other and we infer that the clustering observed in the images of Figure (\ref{fig20}) can be reproduced by a bit larger $\epsilon$ or $n$ than the one that we are using in our model.
Thus, we conclude that our NN performs well on actual N-body simulations which have approximately the same amount of clustering with our training data, and it is able to identify the correct artificial lower cutoff of the halo mass function.

%Moreover, the negative correlation between the number of halos, $N$, and the $\log_{10}$ cutoff, c, that we expect to have because of the shape of the halo mass function (in the simulations there are more haloes with small masses), can be seen in Fig. (\ref{cont2}).
Finally, we can generate mock images which have the same number of haloes and cutoff with N-body simulation images, and compare them in pixel space. More details about this can be found in \aref{comp1}.

%\begin{figure*}
%\centering
%\hspace{-1cm}
%\begin{minipage}{.5\columnwidth}
%\subfloat[\label{17-0} N-body simulation image]{\includegraphics[scale=.4]{images/03.png}}
%\end{minipage} 
%\hspace{2cm}
%\begin{minipage}{.5\columnwidth}
%\centering
%\subfloat[\label{17-1} mock %image]{\includegraphics[scale=.4]{images/imgmock1.png}}
%\end{minipage}\par\medskip
%\centering
%\subfloat[Comparison of \ref{17-0} and \ref{17-1} in pixel space.]{\includegraphics[scale=.45]{images/gistmock1.png}}

%\caption{Comparison of a mock and an N-body simulation image which have the same number, $N$, and the same inner slope, $a$ of the halo mass function.}
%\label{fig17}
%\end{figure*}

%\begin{figure*}
%\begin{minipage}{0.33\textwidth}
%\subfloat[\label {20-0}]{\includegraphics[scale=.4]{images/8_1.png}}
%\end{minipage}%
%\hspace{0.1cm}
%\begin{minipage}{0.33\textwidth}
%\subfloat[\label {20-1}]{\includegraphics[scale=.4]{images/8_2.png}}
%\end{minipage}
%\hspace{0.3cm}
%\begin{minipage}{0.33\textwidth}
%\subfloat[\label {20-2}]{\includegraphics[scale=.4]{images/8_4.png}} 
%\end{minipage}
%\begin{center}
%    This is a mid-figure caption 
%\end{center}
%\caption{Like Figure(\ref{fig1}), but the haloes have mass %$\text{M}_{200\rho_{crit}} \in (10^{8.75},10^{12}) \text{M}_{\odot} \text{h}^{-1}$ and the images are generated from different 12.5x12.5x10.5 $\text{Mpc}^{3}$ sub-boxes of simulation L025N0376.}
%\label{fig20}
%\end{figure*}

%\begin{figure*}
%\begin{minipage}{0.33\textwidth}
%\subfloat[\label {21-0}]{\includegraphics[scale=.4]{images/10_1.png}}
%\end{minipage}%
%\hspace{0.1cm}
%\begin{minipage}{0.33\textwidth}
%\subfloat[\label {21-1}]{\includegraphics[scale=.4]{images/10_2.png}}
%\end{minipage}
%\hspace{0.3cm}
%\begin{minipage}{0.33\textwidth}
%\subfloat[\label {21-2}]{\includegraphics[scale=.4]{images/10_4.png}}
%\end{minipage}
%\caption{Like Figure(\ref{fig1}), but the haloes have mass $\text{M}_{200\rho_{crit}} \in (10^{9.5},10^{12}) \text{M}_{\odot} \text{h}^{-1}$ and the images are generated from different 12.5x12.5x25 $\text{Mpc}^{3}$ sub-boxes of simulation L025N0376.}
%\label{fig21}
%\end{figure*}

\section{Conclusions}

\label{conclusion}
In this work, we have shown how a simple analytical toy halo model can be used together with the likelihood-free inference technique 
Marginal Neural Ratio Estimation (MNRE), to infer parameters of dark matter-only cosmological simulations. 
Our toy model consists of a large number of parameters related to the halo mass function and to the spatial distribution of dark matter haloes.  Such spatial distribution is generated from 2D realizations of Gaussian random fields determined by a power-law power spectrum.
Most of the parameters of the model are nuisance latent variables that correspond to random samples from various distributions, while only four are kept explicitly: the number, $N$, of the dark matter haloes, the slope, $a$, of the halo mass function and the parameters $n,\epsilon$ of our effective clustering model.

By applying MNRE\footnote{through the open-source code \textit{swyft}}, we trained several neural networks with mock images generated by our model, to identify the correct values of the physical parameters that produced a given image. 
We showed that by using the trained neural networks on dark-matter-only simulation images and rotations of them, it is possible to achieve several key goals. 

First, as illustrated in \autoref{fig12}, we managed to reconstruct the halo mass function of those images, i.e., the number, $N$, of the haloes and the inner slope, $a$, of the halo mass function, reasonably well, as the correct values of those parameters always lie inside the $1\sigma$ and $2\sigma$ intervals.  Moreover, as shown in \autoref{fig22}, we were able to identify correctly the lowest mass of the haloes, i.e., the $\log_{10}$ cutoff, $c$, of the halo mass function, while there were some deficiencies in reconstructing  the number of haloes, especially when looking at images where the haloes were more clustered than those in our training data (see Box 2 of \autoref{fig22} and \autoref{lowercutoff} for details). This behaviour is related to the fact that the clustering model that we use is ad-hoc and indicates a direction of future improvement. 
Furthermore, we observed the surprising fact that the results obtained by each image and its three rotations in both \autoref{fig12} and \autoref{fig22} are very stable, despite the fact that rotations change the images completely.
Finally, as shown in \autoref{fig18} and \autoref{fig27}, we were able to generate mock images which look similar to dark-matter-only simulations .

Despite the fact that our results are promising, we note that our model is not perfect and certainly, there are aspects of it that need improvement and refinement.
As mentioned above,  the model that we use to describe the clustering of the haloes, is an effective model, and apart from the fact that is physically motivated, it does not account for higher order correlations between the haloes. Therefore, we restricted our analysis to massive haloes, i.e., to haloes whose clustering can be approximated by a Gaussian field with a power-law power spectrum. However, in order to have a model which is completely physical and can also be used to produce filamentary structures, non-Gaussian features should be introduced. This is a non-trivial task, but previous works in which the nonlinear large-scale structure of the Universe is predicted by deep neural networks using the Zel’dovich approximation (\citealt{2019PNAS..11613825H}) or even using normalizing flows (\citealt{2021arXiv210512024R}) could be used.
Another direction of future improvement could be that of the refinement of the algorithm defining our simulator, by finding and easing its bottlenecks. Now, approximately one second is needed to construct an image with 2000 haloes. That means that generating thousands of mock images is feasible but time consuming. Moreover, using a more complicated model for clustering as suggested above, will probably result in a slowdown of the simulator. Therefore, attention should be given at the optimization of our code.

\newpage

\section*{Acknowledgements}
This work is part of a project that has received funding from the European Research Council (ERC) under the European Union’s Horizon 2020 research and innovation programme (Grant agreement No. 864035). CAC acknowledges the support by the Dutch Research Council (NWO Veni 192.020).
This work was carried out on the Dutch national e-infrastructures, Snellius and Lisa Cluster, with the support of SURF Cooperative. We acknowledge the use of the Python (\citealt{10.5555/1593511}) modules, matplotlib (\citealt{Hunter:2007}), NumPy (\citealt{harris2020array}), SciPy (\citealt{2020SciPy-NMeth}), PyTorch (\citealt{NEURIPS2019_9015}), tqdm (\citealt{casper_da_costa_luis_2021_4663456}),
and Jupyter (\citealt{Kluyver2016jupyter}).

%%%%%%%%%%%%%%%%%%%%%%%%%%%%%%%%%%%%%%%%%%%%%%%%%%
\section*{Data Availability}

The EAGLE data used in this article is available at \url{http://virgodb.dur.ac.uk/} (\citealt{2016A&C....15...72M}; \citealt{2017arXiv170609899T}).
%The data underlying this article will be shared on reasonable request to the corresponding author.


\bibliographystyle{mnras}
\bibliography{example} % if your bibtex file is called example.bib

%\begin{thebibliography}{99}
%\end{thebibliography}

%%%%%%%%%%%%%%%%%%%%%%%%%%%%%%%%%%%%%%%%%%%%%%%%%%

%%%%%%%%%%%%%%%%% APPENDICES %%%%%%%%%%%%%%%%%%%%%

\appendix
\section{Sampling the halo mass function}
\label{sampling}
The sampling can be described in two steps.
 First, we convert the halo mass function into a probability density function (PDF). We want to sample haloes with masses in the range $(10^{9},10^{12})$ ${\text{M}_\odot}$, since this is also the mass range which corresponds to the haloes in the images of \autoref{fig1}.
Therefore, the PDF of the halo mass function can be calculated as

\begin{align}
    \text{PDF}(M)=\frac{b{\left(\frac{M}{{\text{M}_\odot}}\right)}^{-a}}{\int_{10^{9}}^{10^{12}}b\left(\frac{M}{{\text{M}_\odot}}\right)^{-a}d\left(\frac{M}{{\text{M}_\odot}}\right)}=\frac{ (1-a)}{(10^{12})^{1-a}-(10^9)^{1-a}}\cdot \left(\frac{M}{{\text{M}_\odot}}\right)^{-a}.
    \label{PDF}
\end{align}

Then,  we sample from this PDF by using inverse transform sampling (\citealt{devroye:1986}). 
The cumulative distribution function of the PDF of the halo mass function, can be calculated as

\begin{align}
    F(M)&=\frac{\int_{10^{9}}^{M} \left(\frac{M'}{{\text{M}_\odot}}\right)^{-a}d\left(\frac{M'}{{\text{M}_\odot}}\right)}{\int_{10^{9}}^{10^{12}} \left(\frac{M'}{{\text{M}_\odot}}\right)^{-a}d\left(\frac{M'}{{\text{M}_\odot}}\right)}= \\
    &=\frac{\left(\frac{M}{{\text{M}_\odot}}\right)^{1-a}-(10^9)^{1-a}}{(10^{12})^{1-a}-(10^9)^{1-a}},
\end{align}
for $ M \in (10^{9},10^{12})$ $\text{M}_{\odot}$. Solving the equation
\begin{align}
    y= \frac{(\frac{M}{{\text{M}_\odot}})^{1-a}-(10^9)^{1-a}}{(10^{12})^{1-a}-(10^9)^{1-a}}
\end{align}
for $\frac{M}{{\text{M}_\odot}}$ in terms of $y$, yields
\begin{align}
  F^{-1}(y)=(y\cdot(10^{12})^{1-a}-y\cdot(10^{9})^{1-a}+(10^{9})^{1-a})^{\frac{1}{1-a}}.
\end{align}
Therefore, the samples are the values of function $F^{-1}(y)$ for $y \in(0,1).$

As we can see from \autoref{PDF}, the sampling procedure does not depend on the normalization, $b$, of the halo mass function. Thus, $b$ is not a parameter of our model. On the contrary, the inner slope of the halo mass function, $a$, is the second parameter of our model.
Nevertheless, the value of $b$ can be calculated from \autoref{halomassfunction} along with the number, N, of haloes as
 
 \begin{align}
     \frac{dn}{dM}&=b\left(\frac{M}{{\text{M}_\odot}}\right)^{-a}\Rightarrow \nonumber \\
     \int_{N_{1}}^{N_{2}} dn&=\int_{10^{9}}^{10^{12}}b\left(\frac{M}{{\text{M}_\odot}}\right)^{-a}dM \Rightarrow \nonumber  \\
    \int_{N_{1}}^{N_{2}} dn&=\int_{10^{9}}^{10^{12}}b\left(\frac{M}{{\text{M}_\odot}}\right)^{-a}d\left(\frac{M \text{M}_{\odot}}{{\text{M}_\odot}}\right) \Rightarrow \nonumber  \\
\frac{N_{2}-N_{1}}{V}&=\frac{N}{V}=b\left(\frac{(10^{12})^{1-a}}{1-a}-\frac{(10^{9})^{1-a}}{1-a}\right) \text{M}_{\odot} \Rightarrow \nonumber  \\
b&=(1-a)\cdot \frac{N}{((10^{12})^{1-a}-(10^{9})^{1-a})V \text{M}_{\odot}}
 \end{align}
Thus
\begin{align}
    b=(1-a)\cdot \frac{N}{((10^{12})^{1-a}-(10^{9})^{1-a})V}[{\rm \text{Mpc}}^{-3}\text{M}_\odot^{-1}].
\end{align}

\section{Effective Clustering Model}
\label{realizations}
\subsection{Realizations of Gaussian Fields on an $M\times M$ grid}
\label{method_clustering}
To generate realizations of a Gaussian perturbation field, $\delta$, which has a power spectrum given by a power-law, i.e., $P(k)= k^{-n}$, we will first generate a field $\varphi_{\textbf{k}}$ which has a unit amplitude, i.e., $\left\langle \varphi_{\textbf{k}}\varphi_{\textbf{-k}}\right\rangle=1$. Then, we can create the field, $\delta$, which has the desired power spectrum, $P(k)$, by defining $\delta_{\textbf{k}}\equiv P^{1/2}(k)\varphi_{\textbf{k}}$ since
\begin{align}
 \left\langle \delta_{\textbf{k}}\delta_{\textbf{-k}}\right\rangle=P(k) \left\langle \varphi_{\textbf{k}}\varphi_{\textbf{-k}}\right\rangle=P(k).
 \label{eq-35}
 \end{align}
A field which has a constant power spectrum is a white noise field for which:
\begin{align}
    \left\langle \varphi(\textbf{x})\varphi(\textbf{y})\right\rangle &= A\delta^{d}(\textbf{x}-\textbf{y}) \text{ and} \nonumber \\ 
    P(k)&= A ,
    \label{eq-36}
    \end{align}
i.e., for which there are no correlations between the fluctuations at different points.
Therefore, to generate a field, $\varphi_{\textbf{k}}$, that has a unit amplitude, we pick A=1 and we get
\begin{align}
\left\langle \varphi(\textbf{x})\varphi(\textbf{y})\right\rangle = \delta^{d}(\textbf{x}-\textbf{y}).
\label{eq-37}
\end{align}

To create a realisation of white noise on an $M \times $ grid -where $M$ is an even number- we first replace the field $\varphi(\textbf{x})$ with discrete values $\varphi^{\textbf{x}}_{ab}$, where a,b $\in$ \{0,1,...,$M-1$\}, which are drawn from the probability distribution function

\begin{align}
 \mathcal{P}[\varphi^{\textbf{x}}_{ab}]=\prod_{c,d=0}^{M-1}\frac{\exp\left[-\frac{1}{2}\left(\varphi^{\textbf{x}}_{cd}\right)^2\right]}{\sqrt{2\pi}}.
\label{eq-38}
\end{align}
Therefore, to generate a white noise realisation we choose randomly, for every grid point, a value from a standard normal distribution.

Now, we will work in Fourier space, since we want to generate a field that has a desired power spectrum.
First, we Fourier transform the white noise realization. 
On the grid, the continuous Fourier transform that relates $\phi(\mathbf{x})$ and $\varphi_{\mathbf{k}}$ is replaced by the discrete Fourier transform

\begin{align}
\varphi^{\textbf{k}}_{ab}&=\sum^{M-1}_{c,d=0}\exp(-ix_{c}k_{a}-ix_{d}k_{b})\varphi^{\textbf{x}}_{cd} \nonumber \\
\varphi^{\textbf{x}}_{ab}&=\frac{1}{M^{2}}\sum^{M-1}_{c,d=0}\exp(ix_{c}k_{a}+ix_{d}k_{b})\varphi^{\textbf{k}}_{cd} ,
\label{eq-39}
\end{align}
where $x_{a}=a$ and $k_{a}=\frac{2\pi a}{M}$.
We also know that $\varphi^{\textbf{x}}_{ab}$ is real so we have
\begin{align}
\varphi^{*\textbf{x}}_{ab}=\varphi^{\textbf{x}}_{ab} \Rightarrow \varphi^{*\textbf{k}}_{ab}= \varphi^{\textbf{k}}_{-a-b}.
\label{eq-40}
\end{align}
Moreover, \autoref{eq-39} can also be evaluated outside of the domain a,b $\in$ \{0,1,...,$M-1$\} and the extended sequence is $M$-periodic.
Thus, we have 
\begin{align}
\varphi^{\textbf{x}}_{ab}=\varphi^{\textbf{x}}_{(a+nM)b}=\varphi^{\textbf{x}}_{a(b+nM)} \text{ where } n\in \mathbb{Z}
\label{eq-41}
\end{align}
and we also get 
\begin{align}
    \varphi^{*\textbf{k}}_{a(\frac{M}{2}+b)}=\varphi^{\textbf{k}}_{-a-(\frac{M}{2}+b)}=\varphi^{\textbf{k}}_{M-a(-\frac{M}{2}-b+M)}=\varphi^{\textbf{k}}_{(M-a)(\frac{M}{2}-b)}
    \label{eq-42}
\end{align}
and
\begin{align}
    \varphi^{*\textbf{k}}_{(\frac{M}{2}+a)b}=\varphi^{\textbf{k}}_{-(\frac{M}{2}+a)-b}=\varphi^{\textbf{k}}_{(-\frac{M}{2}-a+M)(M-b)}=\varphi^{\textbf{k}}_{(\frac{M}{2}-a)(M-b)}.
    \label{eq-43}
\end{align}

Now, as we saw before, we have to multiply the Fourier space realisation of the white noise with the square root of the desired power spectrum, and the resulting field has to obey \autoref{eq-40} and \autoref{eq-41}, as well as \autoref{eq-42} and \autoref{eq-43}. 
If we would just multiply the Fourier space grid realization of the white noise with the square root of the power spectrum evaluated at points $k$,
where $k=\frac{2\pi}{M}\sqrt{(a^{2}+b^{2})}$ with a,b $\in$ \{0,1,...,$M$-1\}, then we would find
\begin{align}
      \delta^{*\textbf{k}}_{a(\frac{M}{2}+b)}\neq\delta^{\textbf{k}}_{(M-a)(\frac{M}{2}-b)} 
          \label{eq-44}
\end{align}
and
\begin{align}
 \delta^{*\textbf{k}}_{(\frac{M}{2}+a)b}\neq \delta^{\textbf{k}}_{(\frac{M}{2}-a)(M-b)},
     \label{eq-45}
\end{align}
because of the power spectrum factor. If we would then transform the resulting field back to the position space, we would have an imaginary field $\delta^{\textbf{x}}_{ab}$. 
One way to overcome this problem is to define the field $\delta^{\textbf{k}}_{ab}$ as

\vspace{0.2cm}
\noindent  $\delta^{\textbf{k}}_{ab}$
\begin{equation} 
 \resizebox{\columnwidth}{!}{$=\begin{cases}
                P^{1/2}(k)\varphi^{\textbf{k}}_{ab} &  k=\frac{2\pi}{M}\sqrt{a^{2}+b^{2}} \text{ and }a,b\le M/2\\
                P^{1/2}(k)\varphi^{\textbf{k}}_{a(b-M)} & k=\frac{2\pi}{M}\sqrt{a^{2}+(b-M)^{2}} \text{ and } a\le M/2, b>M/2\\
                P^{1/2}(k)\varphi^{\textbf{k}}_{(a-M)b} & k=\frac{2\pi}{M}\sqrt{(a-M)^{2}+b^{2}} \text{ and } a> M/2, b\le M/2\\
                P^{1/2}(k)\varphi^{\textbf{k}}_{(a-M)(b-M)} & k=\frac{2\pi}{M}\sqrt{(a-M)^{2}+(b-M)^{2}} \text{ and } a,b> M/2.
                \end{cases}$ }
\label{eq-46}
\end{equation}
Because of \autoref{eq-41} we can also write

\vspace{0.2cm}
\noindent  $  \delta^{\textbf{k}}_{ab}$
\begin{equation} 
 \resizebox{\columnwidth}{!}{$=\begin{cases}
                P^{1/2}(k)\varphi^{\textbf{k}}_{ab} &  k=\frac{2\pi}{M}\sqrt{a^{2}+b^{2}}  \text{ and }a,b\le M/2\\
                P^{1/2}(k)\varphi^{\textbf{k}}_{ab} & k=\frac{2\pi}{M}\sqrt{a^{2}+(b-M)^{2}} \text{ and } a\le M/2, b>M/2\\
                P^{1/2}(k)\varphi^{\textbf{k}}_{ab} & k=\frac{2\pi}{M}\sqrt{(a-M)^{2}+b^{2}} \text{ and } a> M/2, b\le M/2\\
                P^{1/2}(k)\varphi^{\textbf{k}}_{ab} & k=\frac{2\pi}{M}\sqrt{(a-M)^{2}+(b-M)^{2}} \text{ and } a,b> M/2
                \end{cases}  $ }
  \label{eq-4777}\end{equation}
or, alternatively, we can define

\begin{align}
    \delta^{\textbf{k}}_{(a+M)(b+M)}=\delta^{\textbf{k}}_{ab}=P^{1/2}(k)\varphi^{\textbf{k}}_{ab}=P^{1/2}(k)\varphi^{\textbf{k}}_{(a+M)(b+M)},
     \label{eq-47}
\end{align} where 
 $k=\frac{2\pi}{M}\sqrt{a'^{2}+b'^{2}}$, with $a',b'$ $\in$ \{0,...,$M/2$,$-M/2$+1,..,-1\}.
The real, position space field then reads
\begin{align}
    \delta^{\textbf{x}}_{ab}=\frac{1}{M^{2}}\sum^{M-1}_{c,d=0}\exp(ix_{c}k_{a}+ix_{d}k_{b})\delta^{\textbf{k}}_{cd}. 
         \label{eq-48}
\end{align}

Equivalently, to generate  2D realizations of Gaussian fields, one could use available Python packages, e.g., \textit{powerbox} (\citealt{Murray2018}).

\subsection{Application to mock data generation}
\label{mock data application}
\begin{figure*}

\begin{minipage}{0.33\textwidth}
\vspace*{1cm}
\subfloat[\label {2-1}]{\includegraphics[scale=.4]{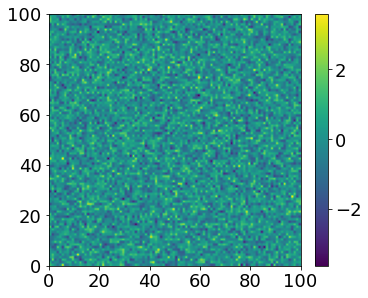}}
\end{minipage}%
\begin{minipage}{0.33\textwidth}
\vspace*{1cm}
\subfloat[\label {2-2}]{\includegraphics[scale=.4]{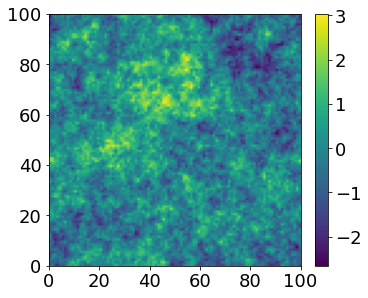}}
\end{minipage}
\begin{minipage}{0.33\textwidth}
\vspace*{1cm}
\subfloat[\label {2-3}]{\includegraphics[scale=.4]{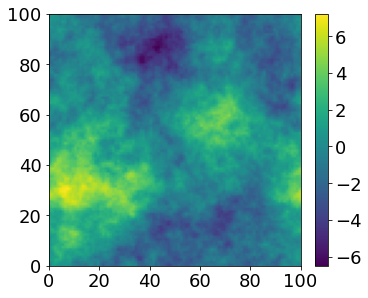}}
\end{minipage}
\caption{Realizations of power spectra of the form $P(k)=k^{-n}$ on three $100 \times 100$ grids. Each grid corresponds to a different slope $n$ with $n\in$ \{0,2,4\}. The $n=0$ realization (\autoref{2-1}) is white noise, while the length scale over which fluctuations are correlated increases as $n$ increases.}
\label{fig2}
\end{figure*}

To generate realizations of the Gaussian fields and thus to add clustering to our model, we follow the procedure described in \aref{method_clustering}.
First, we start by generating a position space realization of a white noise field $\varphi^{\textbf{x}}_{ab}$ with unit amplitude, on a $M \times M$ grid, i.e., $a,b \in \{0,..,M-1\}$. Since images in \autoref{fig1} are generated by 2D histograms with 100 bins, we pick $M$=100.  We call this grid, the “sampling grid'', in order to avoid confusion with other grids that will be defined.
Afterwards, we Fourier transform the white noise realization ($\varphi^{\textbf{x}}_{ab} \rightarrow \varphi^{\textbf{k}}_{ab}$) and we multiply it with the square root of the power spectrum, $P(k)$ of \autoref{powersp}. %,  evaluated at all points ($k_{a},k_{b}) $ $\in$ \{0,...M/2,-M/2+1,..,-1\} with $k=\frac{2\pi}{N}\sqrt{k_{a}^{2}+k_{b}^{2}}$. 
Finally, we perform an inverse Fourier transform to return back to position space and get the desired realization, $\delta^{\textbf{x}}_{ab}$ .
\autoref{fig2}) illustrates 2D realizations of Gaussian fields generated from power spectra with different slopes.  
 
\subsection{Sampling positions of haloes}
\label{details position}

To obtain the positions of the haloes, we randomly sample from the field $f$ and the samples that we obtain are $i$ and $j$ indices of the $100 \times 100$ “sampling grid'' (see \aref{mock data application}). Moreover, we define a new $100\times 100$ grid, called the “coordinate grid'', whose values correspond to pairs of $x$ and $y$ coordinates where ($x,y$) $\in$ (0,12.5) Mpc. Then, the positions $X$ and $Y$ of the haloes are given by the values $x$ and $y$ of the “coordinate'' grid, at the indices that we previously sampled from the “sampling grid''.

Once we obtain the positions of the haloes, we place each one of them in the 2D-sky. This process is accomplished in four steps. We start by subtracting the coordinates of each of the $N$ haloes (as pairs of $X,Y$ values) from the values of the “coordinate grid''. In this way, we construct $N$ $100\times 100$ grids. Then, for each of these grids, we calculate the root sum square of the two values in each one of its cells. If we think of the grids as images and their cells as pixels, this means that the value of each pixel corresponds to the distance between the center of the pixel and a fixed reference, which is the position of the halo. To further clarify that, in each grid we calculate the projected radius, $r'$, of each halo.
  
\section{Simulation based inference}
\subsection{Loss function}
\label{proof}
Following  \cite{hermans2020likelihoodfree}, the loss functional

\vspace{0.2cm}
\noindent $ L[d_{\phi}(\boldsymbol{x},\boldsymbol{\theta})]$
\begin{equation}
   \resizebox{\columnwidth}{!}{$=\int d\boldsymbol{\theta} \int d\boldsymbol{x} p(\boldsymbol{x},\boldsymbol{\theta})[-\log d_{\phi}(\boldsymbol{x},\boldsymbol{\theta})]+p(\boldsymbol{\theta})p(\boldsymbol{x})[-\log(1-d_{\phi}(\boldsymbol{x},\boldsymbol{\theta}))]$}
\end{equation}
is minimized for a function $d^{*}(\boldsymbol{x},\boldsymbol{\theta})$ such that

\begin{multline}
\frac{\delta }{\delta \phi} \left.L[d_{\phi}(\boldsymbol{x},\boldsymbol{\theta})]\right|_{d^{*}}=0  \Rightarrow \\
 \resizebox{\columnwidth}{!}{$-\frac{\delta }{\delta \phi}\int d\boldsymbol{\theta} \int d\boldsymbol{x} p(\boldsymbol{x},\boldsymbol{\theta})[\log d_{\phi}(\boldsymbol{x},\boldsymbol{\theta})]+p(\boldsymbol{\theta})p(\boldsymbol{x})[\log(1-d_{\phi}(\boldsymbol{x},\boldsymbol{\theta}))]=0.$}
\end{multline}
This equation holds for any  $\frac{\delta d_{\phi}(\boldsymbol{x},\boldsymbol{\theta})}{\delta\phi} $, so according to the fundamental lemma of calculus of variations we have
\begin{align}
 -\frac{p(\boldsymbol{x},\boldsymbol{\theta})}{d^{*}_{\phi}(\boldsymbol{x},\boldsymbol{\theta})
  }+\frac{p(\boldsymbol{\theta})p(\boldsymbol{x})}{1-d^{*}_{\phi}(\boldsymbol{x},\boldsymbol{\theta})}=0.
  \label{derivative}
\end{align}
As long as $p(\boldsymbol{\theta})>0$, \autoref{derivative} leads to the decision function
\begin{align}
&\frac{p(\boldsymbol{x},\boldsymbol{\theta})}{d^{*}_{\phi}(\boldsymbol{x},\boldsymbol{\theta})
  }=\frac{p(\boldsymbol{x})p(\boldsymbol{\theta})}{1-d^{*}_{\phi}(\boldsymbol{x},\boldsymbol{\theta})} \Leftrightarrow \nonumber \\
  &d^{*}_{\phi}(\boldsymbol{x},\boldsymbol{\theta})=\frac{p(\boldsymbol{x},\boldsymbol{\theta})}{p(\boldsymbol{x},\boldsymbol{\theta})+p(\boldsymbol{x})p(\boldsymbol{\theta})}.
  \end{align}

\subsection{Neural Ratio Estimation (NRE)}
\label{posterior estimation}

Following \cite{hermans2020likelihoodfree}, \autoref{d} can be rewritten as
\begin{align}
    d(\boldsymbol{x},\boldsymbol{\theta})&\approx\frac{p(\boldsymbol{\theta}|\boldsymbol{x})p(\boldsymbol{x})}{p(\boldsymbol{\theta}|\boldsymbol{x})p(\boldsymbol{x})+p(\boldsymbol{x})p(\boldsymbol{\theta})}=\frac{\frac{p(\boldsymbol{\theta}|\boldsymbol{x})}{p(\boldsymbol{\theta})}}{\frac{p(\boldsymbol{\theta}|\boldsymbol{x})}{p(\boldsymbol{\theta})}+1}.
    \label{pre-ratio}
    \end{align}
The ratio 
\begin{align}
    r(\boldsymbol{x},\boldsymbol{\theta})\equiv \frac{p(\boldsymbol{x}|\boldsymbol{\theta})}{p(\boldsymbol{x})}= \frac{p(\boldsymbol{\theta}|\boldsymbol{x})}{p(\boldsymbol{\theta})}
\end{align}
can be then approximated from \autoref{pre-ratio} as
\begin{align}
    d(\boldsymbol{x},\boldsymbol{\theta})\approx\frac{\frac{p(\boldsymbol{\theta}|\boldsymbol{x})}{p(\boldsymbol{\theta})}}{\frac{p(\boldsymbol{\theta}|\boldsymbol{x})}{p(\boldsymbol{\theta})}+1}&=\frac{r(\boldsymbol{x},\boldsymbol{\theta})}{r(\boldsymbol{x},\boldsymbol{\theta})+1} \Leftrightarrow \nonumber \\
    r(\boldsymbol{x},\boldsymbol{\theta})d(\boldsymbol{x},\boldsymbol{\theta})-r(\boldsymbol{x},\boldsymbol{\theta})&\approx d(\boldsymbol{x},\boldsymbol{\theta})\Leftrightarrow \nonumber \\
        r(\boldsymbol{x},\boldsymbol{\theta})&\approx\frac{d(\boldsymbol{x},\boldsymbol{\theta})}{d(\boldsymbol{x},\boldsymbol{\theta})-1}\Leftrightarrow \nonumber \\
         \frac{p(\boldsymbol{\theta}|\boldsymbol{x})}{p(\boldsymbol{\theta})}&\approx\frac{d(\boldsymbol{x},\boldsymbol{\theta})}{d(\boldsymbol{x},\boldsymbol{\theta})-1}.
\end{align}
The approximation of the ratio allows us finally to to estimate directly the posterior $p(\boldsymbol{\theta}|\boldsymbol{x})$ as
\begin{align}
    p(\boldsymbol{\theta}|\boldsymbol{x})\approx \frac{d(\boldsymbol{x},\boldsymbol{\theta})}{d(\boldsymbol{x},\boldsymbol{\theta})-1}p(\boldsymbol{\theta})=\hat{p}(\boldsymbol{\theta}|\boldsymbol{x}).
\end{align}

\subsection{Marginal Neural Ration Estimation (MNRE)}
\label{MNRE_appendix}

As exactly described in \autoref{ratioestimation}, \textit{swyft} trains binary classifiers $d_{k,\phi}(\boldsymbol{x},\boldsymbol{\vartheta}_{k})$ in parallel, where, now, each one depends on the parameters of interest $\boldsymbol{\vartheta}_{k}$, to distinguish jointly drawn from marginally drawn parameter-simulation pairs. 
The ratios of interest $r_{k}(\boldsymbol{x},\boldsymbol{\vartheta}_{k})$ can be then estimated as
\begin{align}
   r_{k}(\boldsymbol{x},\boldsymbol{\vartheta}_{k})\approx \frac{d_{k,\phi}(\boldsymbol{x},\boldsymbol{\vartheta}_{k})}{1-d_{k,\phi}(\boldsymbol{x},\boldsymbol{\vartheta_{k}})}=\hat{r}_{k}(\boldsymbol{x},\boldsymbol{\vartheta}_{k})
   \label{ratioswyft}
\end{align}
and the corresponding posteriors $p(\boldsymbol{\vartheta}_{k}|\boldsymbol{x})$ can be estimated as
\begin{align}
    p(\boldsymbol{\vartheta}_{k}|\boldsymbol{x})\approx r_{k}(\boldsymbol{x},\boldsymbol{\vartheta}_{k})p(\boldsymbol{\vartheta}_{k})=\hat{p}(\boldsymbol{\vartheta}_{k}|\boldsymbol{x}).
\end{align}

Practically, estimating the ratio using \autoref{ratioswyft} is susceptible to numerical errors since the ratio is computed by transforming the sigmoid function \citep{hermans2020likelihoodfree}. Thus, \textit{swyft} uses an modified network architecture which directly outputs the value of the NN $f_{k,\phi}$ \citep{Miller:2021hys}, instead of $\sigma \circ f_{k,\phi}$ which the majority of the binary classifiers output. 
Since $\sigma(x)$ is defined by the formula
\begin{align}
    \sigma(x)=\frac{e^{x}}{1+e^{x}},
\end{align}
\autoref{pre-ratio}, which describes the output of a typical classifier, after being adjusted to marginals, can be written as

\begin{align}
    d_{k,\phi}(\boldsymbol{x},\boldsymbol{\vartheta}_{k})\approx\frac{\frac{p(\boldsymbol{\vartheta}_{k}|\boldsymbol{x})}{p(\boldsymbol{\vartheta}_{k})}}{\frac{p(\boldsymbol{\vartheta}_{k}|x)}{p(\boldsymbol{\vartheta}_{k})}+1}&=\sigma \circ \log(\frac{p(\boldsymbol{\vartheta}_{k}|\boldsymbol{x})}{p(\boldsymbol{\vartheta}_{k})})= \\
    &=\sigma \circ \log(r_{k}(\boldsymbol{x},\boldsymbol{\vartheta}_{k})).
\end{align}
Therefore, the modified classifier implies
\begin{align}
   f_{k,\phi}=\log(r_{k}(\boldsymbol{x},\boldsymbol{\vartheta}_{k})).
\end{align}
Using \autoref{ratiomargswyft} we can then estimate the marginal posteriors as
\begin{align}
    p(\boldsymbol{\vartheta}_{k}|\boldsymbol{x})\approx e^{f_{k,\phi}}p(\boldsymbol{\vartheta}_{k})=\hat{p}(\boldsymbol{\vartheta}_{k}|\boldsymbol{x}).
\end{align}

\onecolumn
\newpage
\vspace*{-22.7cm}
\section{2D realizations of Gaussian fields}

\vspace{6.4cm}

\section{Comparison of N-body simulation and mock images}
\label{comp1}
\begin{figure*}
\hspace{-0.7cm}
\begin{minipage}{.33\linewidth}
\vspace*{0.6cm}
%\centering
\subfloat[\label{18-0} N-body simulation]{\includegraphics[scale=.4]{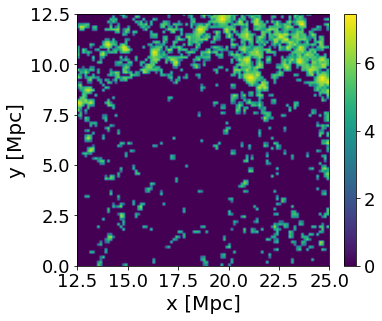}}
\end{minipage}%
%\hspace{-2.5cm}
\begin{minipage}{.33\linewidth}
\vspace*{0.6cm}
%\centering
\subfloat[\label{18-1} mock image]{\includegraphics[scale=.4]{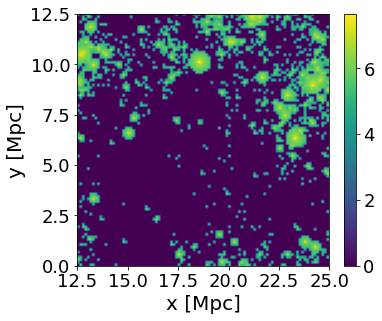}}
\end{minipage}
%\centering
\begin{minipage}{.33\linewidth}
\vspace*{0.6cm}
\subfloat[Comparison of \autoref{18-0} and \ref{18-1} in pixel space. \label{comparison1}]{\includegraphics[scale=.38]{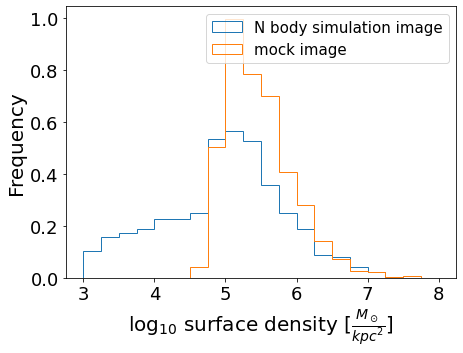}}
\end{minipage}
\caption{Comparison of an N-body simulation and an mock image which have the same number of haloes, $N$, and the same inner slope, $a$ of the halo mass function. The mock image was chosen in order for the clustering of the haloes to visually correspond to that of the N-body simulation image. From the histograms of \autoref{comparison1}, we see that the distributions of the surface densities in pixel space are similar to each other. However, small differences between them are spotted, especially towards low surface densities produced by the boundaries of the haloes. This is unsurprising, since our model is not perfect and, as described in \autoref{toy halo model}, we defined an artificial boundary for the haloes in the mock images.}
\label{fig18}
\end{figure*}

\begin{figure*}
\hspace{-0.7cm}
\begin{minipage}{.33\linewidth}
\vspace*{-0.4cm}
%\centering
%\hspace{-1cm}
\subfloat[\label{27-0} N-body simulation]{\includegraphics[scale=.4]{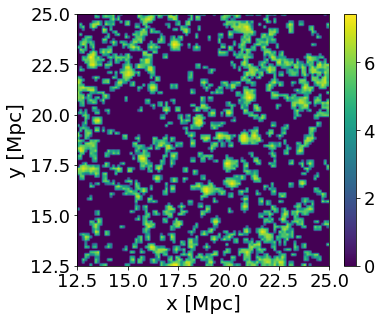}}
\end{minipage}%
%\hspace{-2.5cm}
\begin{minipage}{.33\linewidth}
\vspace*{-0.4cm}

%\centering
\subfloat[\label{27-1} mock image]{\includegraphics[scale=.4]{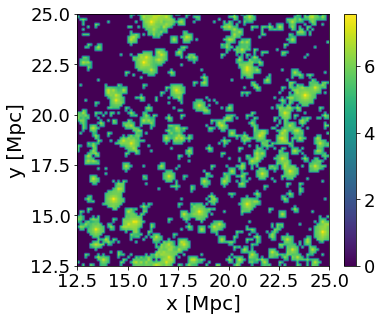}}
\end{minipage}
%\centering
\begin{minipage}{.33\linewidth}
\vspace*{-0.4cm}
\subfloat[Comparison of \autoref{27-0} and \ref{27-1} in pixel space.]{\includegraphics[scale=.38]{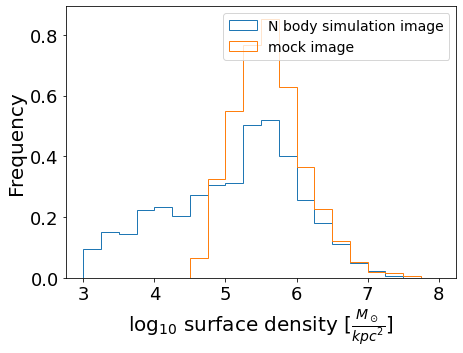}}
\end{minipage}
\caption{Like \autoref{fig18}, but the images are generated using a larger $\log_{10}$ cutoff, $c$, of the halo mass function.}
\label{fig27}
\end{figure*}

%%%%%%%%%%%%%%%%%%%%%%%%%%%%%%%%%%%%%%%%%%%%%%%%%%
\clearpage
\twocolumn

% Don't change these lines
\bsp	% typesetting comment
\label{lastpage}
\end{document}